\documentclass[reprint,amsmath,amssymb,aps,prfluids,onecolumn]{revtex4-2}

\usepackage{graphicx, amsmath, color, epstopdf, epsfig, array, hyperref, mathtools, color, tabularray, booktabs}

\definecolor{White}{rgb}{1.0,1.0,1.0}
\definecolor{Light}{rgb}{0.97,0.97,0.97}
\definecolor{Dark}{rgb}{0.925,0.925,0.925}
\newcommand{\commentcolor}{black}
\newcommand{\commentcolors}{black}

\begin{document}
\title{Reynolds stress decay modeling informed by \\ 
anisotropically forced homogeneous turbulence}
\author{Ty Homan}
    \email{tyhoman@stanford.edu}

\author{Omkar B. Shende}
    \email{oshende@stanford.edu}

\author{Ali Mani}
    \email{alimani@stanford.edu}

\affiliation{Department of Mechanical Engineering, Stanford University, Stanford, CA 94305, USA}

\begin{abstract}
Models for solving the Reynolds-averaged Navier-Stokes equations are popular tools for predicting complex turbulent flows due to their computational affordability and ability to provide or estimate quantities of engineering interest. However, results depend on a proper treatment of unclosed terms, which require progress in the development and assessment of model forms. In this study, we consider the Reynolds stress transport equations as a framework for second-moment turbulence closure modeling. We specifically focus on the terms responsible for decay of the Reynolds stresses, which can be isolated and evaluated separately from other terms in a canonical setup of homogeneous turbulence. We show that by using anisotropic forcing of the momentum equation, we can access states of turbulence traditionally not probed in a triply-periodic domain. The resulting data spans a wide range of anisotropic turbulent behavior in a more comprehensive manner than extant literature. We then consider a variety of model forms for which this data allows us to perform a robust selection of model coefficients and select an optimal model that extends to cubic terms when expressed in terms of the principal coordinate Reynolds stresses. Performance of the selected decay model is then examined relative to the simulation data and popular models from the literature, demonstrating the superior accuracy of the developed model and, in turn, the efficacy of this framework for model selection and tuning. 
\end{abstract}

\maketitle

\section{Introduction} \label{sec:introduction}

Recent advances in computational power have enabled large leaps in the range of scales that can be resolved at reasonable cost in direct numerical simulation (DNS) of turbulent fluid flows. For myriad applications, however, only averaged quantities, such as the mean velocity profile or mean scalar transport, need to be determined; a parsimonious method for computing such quantities of engineering interest is therefore to solve in the averaged space directly. Reynolds averaging, denoted $\overline{\bullet}$, acts in directions of homogeneity and yields the Reynolds-averaged Navier-Stokes (RANS) equations when applied to the governing equations for fluid flow. RANS solvers therefore provide mean quantities in a lower-dimensional space than the full DNS equations. For example, when considering the evolution of an unsteady velocity field, $u_i(\vec{x},t)$, inside a triply periodic domain with homogeneity in all spatial directions, the RANS equations are ordinary differential equations of time only. 

In exchange for this dimensionality reduction, however, the RANS equations have terms which require modeling, and such models require databases of numerical and experimental measurements of realizations of turbulent flow to establish model parameters. While a variety of approaches are used to solve this closure problem, we focus on the use of the six Reynolds stress transport (RST) equations for the second-moment terms unresolved in the RANS equations. The literature on modeling and parameter estimation approaches for the unclosed terms of the RST equations is vast and is documented in many sources, recent examples of which range from textbooks \cite{pope_2000, durbin_2010}, to resources for the wider community \cite{larc}, to reviews \cite{alfonsi_2009, duraisamy_2019, durbin_2018}. 

The aforementioned triply periodic problem setup allows some simplifications of the general RST equation form. In particular, decaying and forced homogeneous turbulence in such a domain form canonical problems for unsteady and statistically stationary flows, respectively. For the case of incompressible, single-component, homogeneous turbulence with zero mean flow, the exact transport equation for the Reynolds stresses, $\overline{u_iu_j}$, reduces to

\begin{equation}  \label{eqn:RS}
\frac{d{\overline{u_iu_j}}}{d{t}} = P_{ij} + \overline{ \frac{p}{\rho} \left(\frac{\partial{u_i}}{\partial x_j} + \frac{\partial u_j}{\partial x_i}\right) } - 2 \nu \overline{\frac{\partial{u_i}}{\partial{x_k}}\frac{\partial u_j}{\partial x_k}} =  P_{ij} + \Pi_{ij} - \varepsilon_{ij} = P_{ij} + D_{ij}
\end{equation}

\noindent where $p$ is the pressure, $\rho$ denotes the fluid density, $\nu$ is the fluid kinematic viscosity, $P_{ij}$ is the production tensor, $\Pi_{ij}$ is the pressure-strain correlation \textcolor{\commentcolor}{given by the middle term,} and $\varepsilon_{ij}$ is the rate of dissipation for each component\textcolor{\commentcolor}{, given by the final term}. The pressure-strain tensor -- also referred to as the \emph{slow pressure-strain} term when a mean velocity gradient is absent \cite{pope_2000}, as in this work -- is traceless in incompressible flows and serves to redistribute energy amongst the six components to promote a return to isotropy, while dissipation removes energy from the flow. The production tensor serves, in total, to add energy to the flow; for simulations of decaying turbulence $P_{ij} = 0$, but in other cases it is often explicitly defined in terms of quantities (\emph{i.e.}, the mean velocity gradients and Reynolds stresses) that do not need to be modeled. 

However, $\Pi_{ij}$ and $\varepsilon_{ij}$ are unclosed terms that require further modeling. In this study we consider these two terms bundled into a single anisotropic term that represents the overall decay of the Reynolds stresses, denoted here as $D_{ij}$ in following similar conventions to \cite{lrr_1975, ssg_1991}. Traditionally, the anisotropy of the dissipation tensor has been considered so negligible that the common ansatz is $\epsilon_{ij} = \frac{2}{3} \epsilon~\delta_{ij}$ \cite{reynolds_1976}. This, however, is not the case \cite{yeung_1991, lumley_1992}.  We choose to model the sum of these two terms so that the finite anisotropy of the dissipation tensor and of the pressure-strain is captured consistently, as has been concluded by others, such as \cite{poroseva_2001}.

Two of the most popular Reynolds stress equation models in the literature and in practial use are the LRR model proposed in \cite{lrr_1975} and the SSG model of \cite{ssg_1991}. These two models have been modified and augmented with additional forms since their proposal -- \emph{e.g.}, at \cite{larc} -- and are complete Reynolds stress models. These models are originally formulated in terms of the traceless Reynolds stress anisotropy tensor, defined as 

\begin{equation}
    b_{ij} = \frac{\overline{u_iu_j}}{2k} - \frac{\delta_{ij}}{3},  
\end{equation}

\noindent where $2k = \overline{u_i u_i}$ is twice the turbulent kinetic energy (TKE) and $\delta_{ij}$ is the Kronecker delta. In this work, the problem setup permits study of the decay term in isolation, so we consider only those corresponding terms in each of these models. 

The LRR model decay term is written as

\begin{equation}  \label{eqn:LR}
D^{\textbf{LRR}}_{ij} = - \frac{C_1}{k/\epsilon} \left(\overline{u_iu_j} - \frac{2}{3}\delta_{ij}k \right) - \frac{2}{3}\epsilon\delta_{ij},
\end{equation}

\noindent where the right-hand-side represents a linear function of the Reynolds stress anisotropy tensor, $\epsilon$ is the scalar rate of dissipation of TKE \textcolor{\commentcolor}{with $2\epsilon = \varepsilon_{ii} = 2\nu \overline{\frac{\partial{u_i}}{\partial{x_k}}\frac{\partial u_i}{\partial x_k}}$}, and $C_1$ is a model constant. \textcolor{\commentcolors}{Its derivation} makes use of constraints imposed by coordinate-system invariance, realizability, dimensional analysis, and tensor properties, such as an appeal to the dissipation tensor being isotropic. The pressure-strain correlation is modeled using a linear return-to-isotropy form from \cite{rotta_1951}, which is a simple representation of the tendency of turbulence towards isotropy as it decays.

The decay term of the SSG model is described in \cite{sarkar_1990}, but we will refer to it as the \emph{SSG model} for clarity. This decay term model is written as

\begin{equation}  \label{eqn:SS}
D^{\textbf{SSG}}_{ij} = -C_1~\epsilon b_{ij} + C_2~\epsilon \left( b_{ik}b_{kj} - \frac{1}{3}b_{mn}b_{nm}\delta_{ij} \right) - \frac{2}{3}\epsilon\delta_{ij}
\end{equation}

\noindent where the right-hand-side is a quadratic function of the Reynolds stress anisotropy tensor and $C_1$ and $C_2$ are model parameters. The SSG model is developed using the same constraints as the LRR model; however, it employs a return-to-isotropy model that is quadratic in the Reynolds stress anisotropy tensor, as first given in \cite{sarkar_1990}. Nonlinear forms have consistenly shown improvement over a linear return-to-isotropy model, so we will focus on such forms here.

Such canonical models for Reynolds stress decay were formulated based on mathematical arguments and, in principle, require a body of experimental and computational data to fit coefficients. As such, the most widely-used coefficients for the LRR and SSG models are specified using analysis of experimental data associated with multiple flow configurations, where each configuration is designed to systematically activate model physics. Probing a wide range of mean flow deformations or flow types therefore requires numerous experimental configurations. Key limitations with both decay term models is that they were developed based on limited datasets of decaying turbulence and that linear and quadratic forms may be unable to capture the higher-order nonlinearities present in the data \cite{chung_1995}.

 Here, we present a method to inform Reynolds stress decay modeling using a more robust selection of data that can be obtained from a single flow configuration in a homogeneous domain. Our method utilizes steady simulations with effective forcing techniques inspired by the linear forcing of \cite{lundgren_2003, rosales_2005}, which has been modified to effectively match the turbulent characteristics of free shear flows in multiple recent works \cite{rah_2018, dhandapani_2019, yi_2023}. The result is an anisotropic forcing term that mimics the effects of turbulent production that can be used to manipulate homogeneous turbulence such that a variety of flows can be effectively probed within a triply periodic domain. This approach offers a unified means to tune and evaluate Reynolds stress decay models with simulations that exhibit stationary statistics in both space and time, and with unsteady simulations of decaying turbulence.
 
 In this work, we probe the Reynolds stress decay term through many independent stationary simulations over a wide range of anisotropic forcing. Section II describes the approach used in our high fidelity simulations and we present the forced simulation results in Section III and their ability to realize turbulent states independent of Reynolds number. Section IV then details the proposed modeling framework, and Section V provides an evaluation of the resulting model against simulations of decaying turbulence and popular models from extant literature, along with further discussion. Overall conclusions are given in Section VI.

\section{Approach}

In this section, we illustrate our  method for approaching anisotropic turbulent flows with zero mean velocity and pressure in a canonical triply periodic domain. This homogeneous anisotropic turbulence (HAT) can be characterized in a two-dimensional diagram, allowing quantification of the subspace of realizable turbulence states.

\subsection{Governing equations} \label{sec:gov-eqns}

Here, we consider a forced incompressible flow governed by

\begin{equation}
    \frac{\partial{u_i}}{\partial{t}} + \frac{\partial{u_i u_j}}{\partial x_j} = - \frac{1}{\rho} \frac{\partial p}{\partial x_i} + \frac{\partial}{\partial x_j} \left( \nu_t \frac{\partial u_i}{\partial x_j} \right)+ \Omega A_{ij}\Tilde{u_j},
    \label{eqn:momentum}
\end{equation}

\noindent in addition to the continuity equation, $\nabla \cdot \mathbf{u} = 0$. In order to represent the limit state of infinite Reynolds number turbulence, we perform large-eddy simulations by employing an eddy viscosity, $\nu_t$, in lieu of molecular viscosity. This means all quantities in Equation \ref{eqn:momentum} are \textit{filtered}, and the filter width is implied to be proportional to the grid size. To capture sub--grid effects, we use a Smagorinsky-Lilly model, where $\nu_t = (C_s\Delta)^2|\overline{S}|$. In accordance with a constant-coefficient model, here $C_s = 0.2$ is a tunable constant consistent with \cite{lilly_1966}, $\Delta$ is the grid size, and $|\overline{S}| = \sqrt{2S_{ij}S_{ij}}$ is the magnitude of the resolved strain-rate tensor. 

The last term in \ref{eqn:momentum} is a linear forcing term which provides turbulent energy production or removal based on elements of the forcing matrix $\boldsymbol{A} = A_{ij}$. \cite{rosales_2005, lundgren_2003} pioneered use of forcing that corresponds to an isotropic $A_{ij} = A\delta_{ij}$, which was generalized by \cite{rah_2018, dhandapani_2019} to a tensor form in order to mimic the Reynolds stress production term associated with canonical shear flows. We build on these works by adding further modifications and by instead considering the inverse problem of finding the realizable states of Reynolds stress that correspond to a freely-chosen $A_{ij}$, thereby representing a range of flow configurations. We then add $\Omega$, a time-varying controller that maintains the TKE at a prescribed level, using a proportional controller framework inspired by \cite{bassenne_2016}. This means we solve with $P_{ij} = \overline{u_i\Omega A_{jk}\Tilde{u_k}} + \overline{u_j\Omega A_{ik}\Tilde{u_k}}$ instead of the standard $P_{ij} = -\overline{u_iu_k}\frac{\partial u_j}{\partial x_k} -\overline{u_ju_k}\frac{\partial u_i}{\partial x_k}$. The forcing matrix therefore plays the effective role of the mean velocity gradient tensor as the means of turbulent energy production.

Furthermore, to minimize sensitivity of the results to domain orientation, the field that multiplies the controller, $\Tilde{u_i}$, is the velocity field passed through a high-pass filter, as in \cite{shende_2024}. The high-pass filter applied to the forced velocity field smoothly varies from 0 at $\kappa=2$ to 1 at $\kappa=3$ using a cosine profile, where $\kappa = \sqrt{\kappa_x^2 + \kappa_y^2 + \kappa_z^2}$ is the magnitude of the wavenumber vector. As a result, energy is only injected at wavenumbers $\kappa \gtrsim 2$. For isotropically forced turbulence, this strategy ensures that a velocity auto-correlation becomes zero within the simulation box, which is not a feature of the standard linear forcing method. As we show in Appendix \ref{app:filter-effects}, the high-pass filter mitigates the dependence of obtained statistics on the orientation of the periodic box in our anisotropically forced simulations.

\begin{figure}[ht]
    \centering
    \includegraphics[width=0.75\textwidth]{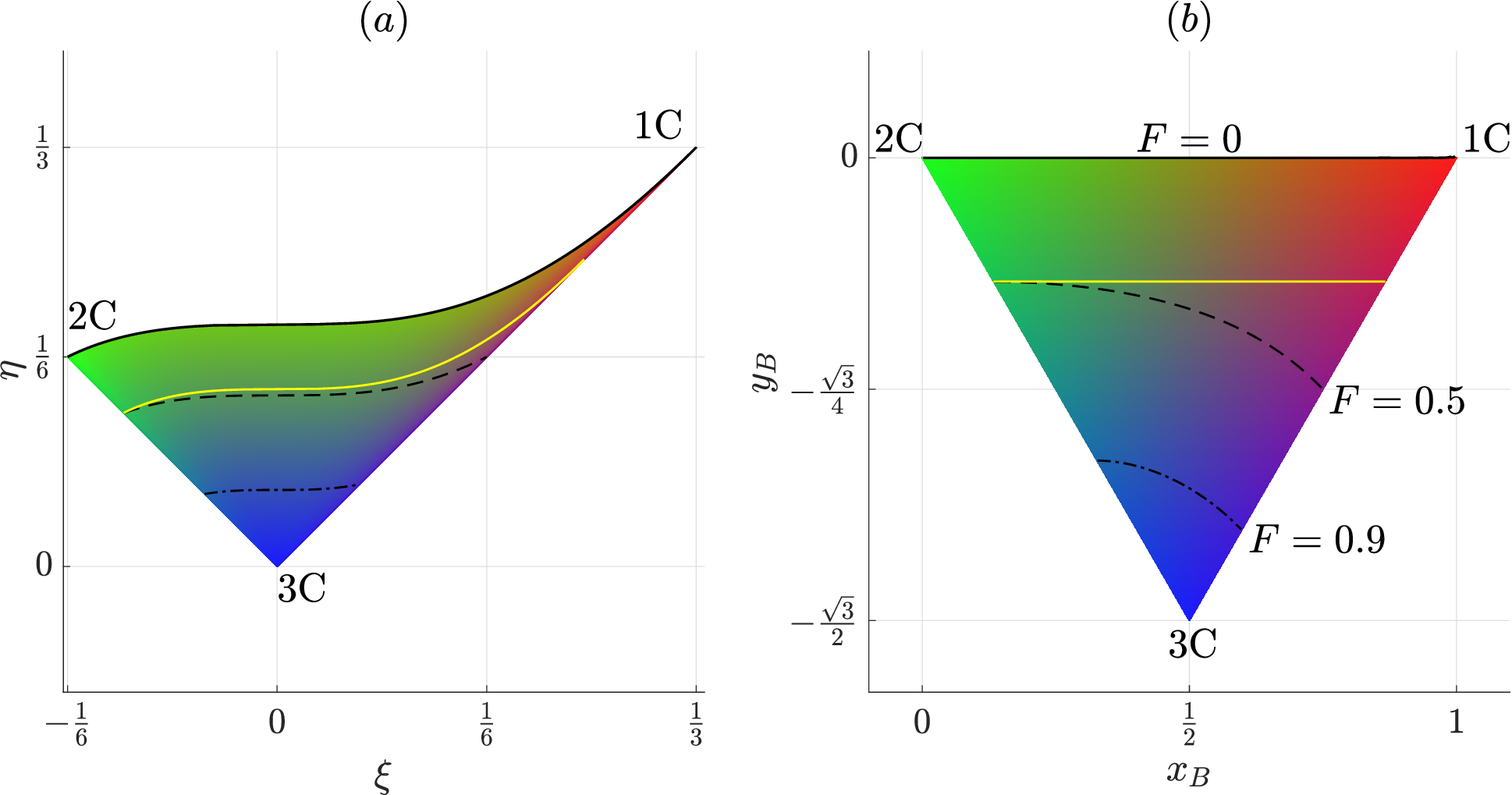}
    \caption{A comparison of the turbulence anisotropy maps used in this paper, with the componentality coloring as suggested in \cite{emory_2014}. Isocontours of $F$ are plotted in black, and an isocontour of $C_3$ as computed in \cite{banerjee_2007} is plotted in yellow. (a) The Lumley triangle is shown in $\xi, \eta$ space and (b) the barycentric triangle is shown with labeled $F$ values.}
    \label{fig:triangles}
\end{figure}

\subsection{Reynolds stress anisotropy}

The controller, $\Omega$, ensures that the TKE is maintained at a constant value, but this means that the Reynolds stress anisotropy tensor, $b_{ij}$, is a function of the anisotropy of the forcing matrix, $A_{ij}$, as it is varied across simulations. The Reynolds stress anisotropy tensor is constrained to be traceless and therefore only has two degrees of freedom, which allows it to be visualized as a two-dimensional surface. Two popular anisotropy invariant mappings used in the literature are the invariant formulation of \cite{choi_lumley_2001}, which is commonly denoted the \emph{Lumley triangle}, and the barycentric triangle developed by \cite{banerjee_2007}. In both of these visualizations, the limiting states of the componentality of $b_{ij}$ (associated with one-, two-, and three-component axisymmetric Reynolds stresses) are represented as vertices of the triangles, as shown in Fig. \ref{fig:triangles}. 

The Lumley triangle uses a domain based on the invariants $\xi$ and $\eta$, where $6\xi^3 = b_{ij}b_{jk}b_{ki}$ and $6\eta^2 = b_{ij}b_{ji}$, so that $\xi$ and $\eta$ are nonlinear functions of the anisotropy tensor eigenvalues. The triangle was created for evaluating Reynolds stress decay trajectories and the nonlinear mapping enables a close visualization near the isotropic corner \cite{choi_lumley_2001}. The barycentric triangle, on the other hand, provides a linear mapping of the eigenvalues of the anisotropy tensor, giving the componentality of $b_{ij}$ equal spatial representation. We employ a version of the barycentric triangle that matches the standard orientation of the Lumley triangle, \emph{viz.} we choose Euclidean coordinates for the limiting states such that $1\text{C} \equiv (1,0)$, $2\text{C} \equiv (0,0)$, and $3\text{C} \equiv (\frac{1}{2},-\frac{\sqrt{3}}{2})$. This can be seen in Fig. \ref{fig:triangles}. 

The role of $F = 1 - 27\eta^2+54\xi^3$ will be discussed in Section III, but here it is key only to note that the mapping between the two triangles distorts the relative size of areas and the relative orientation of lines. For example, the yellow curve, which is a straight line in the barycentric triangle and deviates from the $F = 0.5$ contour significantly along the $1\text{C}-3\text{C}$ leg of the barycentric triangle, does not appear to visually differ as significantly from that same contour on the Lumley triangle. The fact that angles between curves are not preserved demonstrates that the map between triangles, while demonstrated to be bijective by \cite{banerjee_2007}, is not necessarily conformal. 

As each triangle has its advantages, we will use both triangles for plotting decay trajectories and the barycentric triangle for assessing representation of the anisotropy parameter space by forced simulations.

\begin{figure}[hb]
    \centering
    \includegraphics[width=\textwidth]{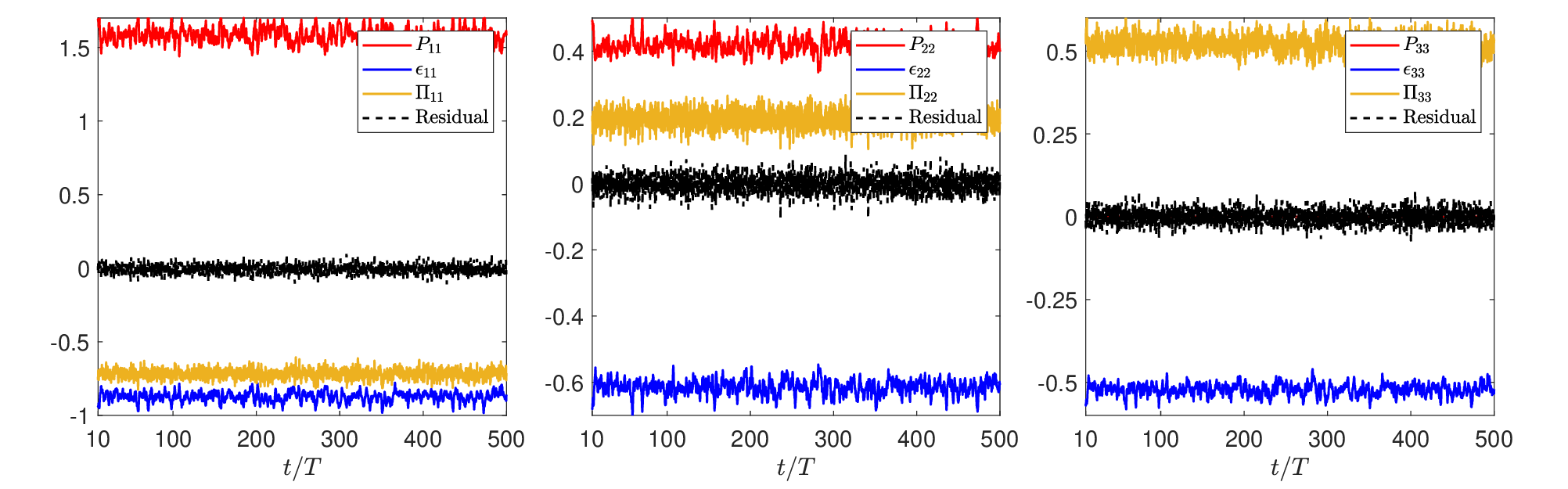}
    \caption{Budgets for the three principal components of Reynolds stress normalized by the mean value of the TKE production, $P_{ii}/2$ show statistical closure. Results correspond to forcing case 14 in Table \ref{tab:allcases}.}
    \label{fig:budget}
\end{figure}

\subsection{Numerical implementation}

\begin{figure}[t]
    \centering
    \includegraphics[width=0.85\textwidth]{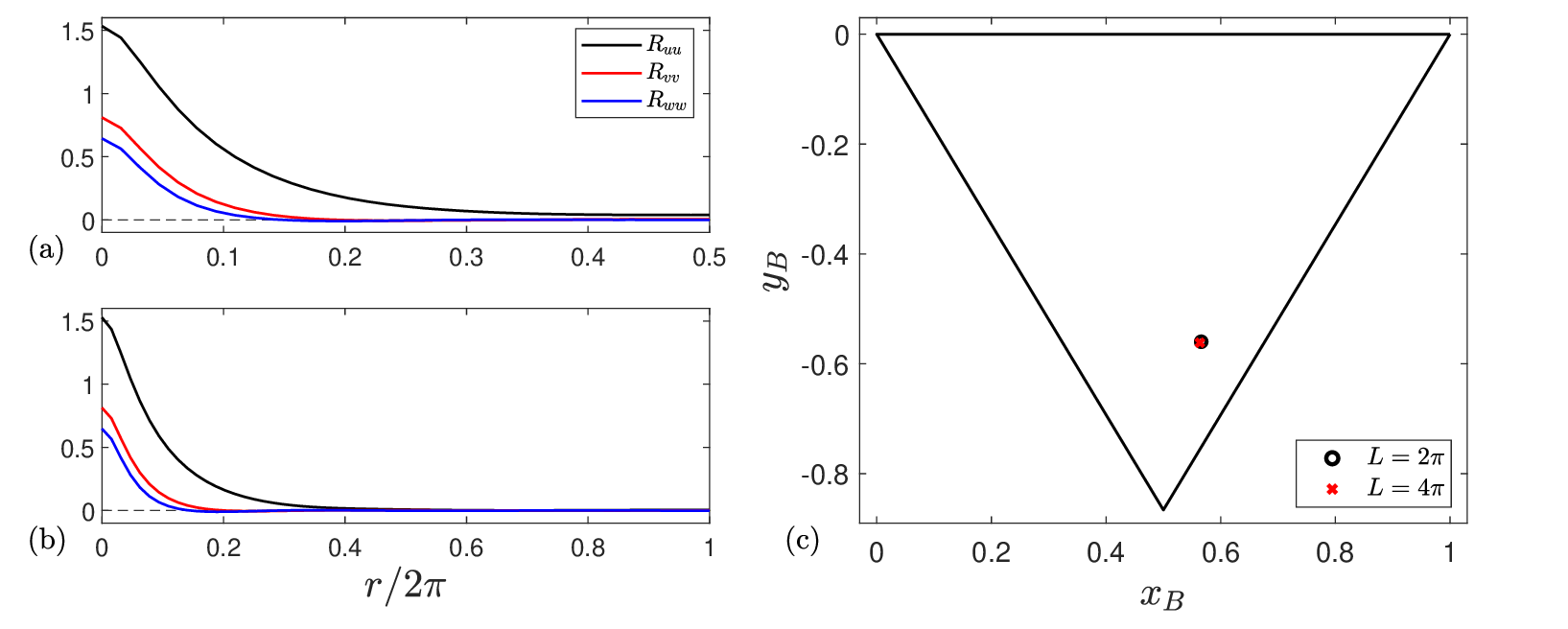}
    \caption{Velocity autocorrelations for forcing case 14 in (a) $L=2\pi$ and (b) $L=4\pi$ domains with matching resolution. The resulting Reynolds stress anisotropies are mapped in (c). $R_{uu}$ is the $u_1$ autocorrelation as a function of $x_1$ displacement, while $R_{vv}$ and $R_{ww}$ are the autocorrelations of $u_2$ and $u_3$ with $x_2$ and $x_3$ displacements, respectively.}
    \label{fig:auto-corr}
\end{figure}

\begin{figure}[b]
    \centering
    \includegraphics[width=0.5\textwidth]{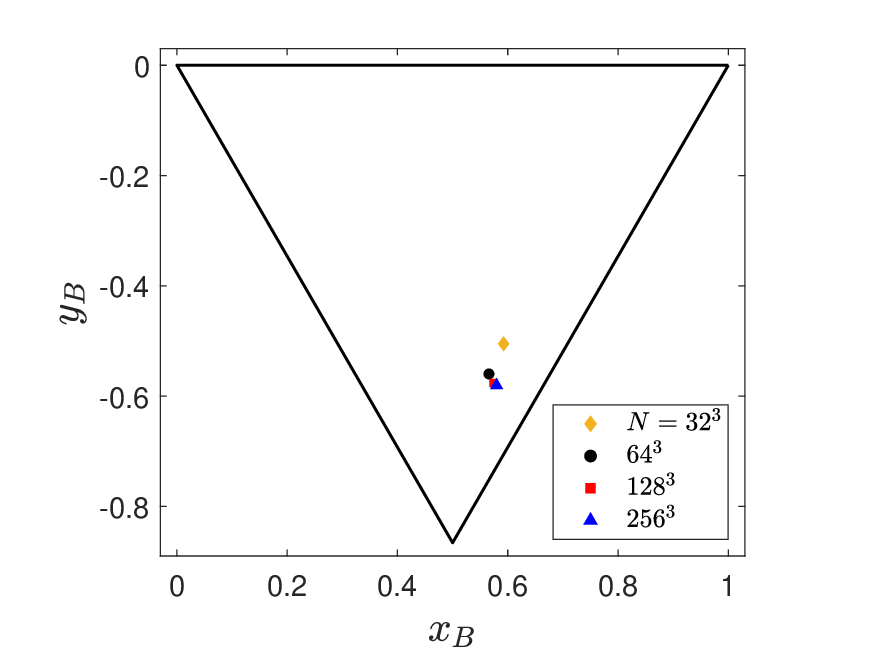}
    \caption{Reynolds stress anisotropies resulting from high-fidelity simulations with various grid resolutions mapped on the barycentric triangle show convergence. Results correspond to forcing case 14.}
    \label{fig:refine}
\end{figure}

The data for this work comes from solving Eqn. \ref{eqn:momentum} on a uniform isotropic grid using a pseudospectral code adapted from the finite volume code of \cite{pouransari_2016}. The code uses Fourier spectral derivatives with explicit time advancement through a fourth-order Runge-Kutta method. Convective terms were dealiased using zero-padding. The spectral derivatives were verified against analytical solutions and the RK4 method provides fourth-order-accurate convergence for decaying turbulence. Further validation is provided by confirming the closure of Reynolds stress budgets as shown in Fig. \ref{fig:budget}. In this plot and later results, time is non-dimensionalized by $T = \frac{L}{2\pi u_{rms}}$. Simulations are performed in a triply periodic box with nominal edge length $L=2\pi$, and $T = 1$ as the controller maintains $u_{rms} = \sqrt{\frac{2k}{3}} $ at 1. The eddy turnover time, computed as the two-sided integral of the spatial velocity autocorrelation normalized by $u_{rms}^3$, was measured to be $0.8 T$ for an isotropically forced simulation.

Forcing cases are constructed using different specifications for the forcing matrix $A_{ij}$, with elements serving to excite or suppress components of the Reynolds stress tensor. For some anisotropic forcings, the high-pass filter is unable to reduce velocity auto-correlations to zero, as shown in Figure \ref{fig:auto-corr}. To investigate the effect of this auto-correlation on the Reynolds stresses, we perform a simulation with matching grid resolution, but with the domain size increased so that $L = 4\pi$. The results in Fig. \ref{fig:auto-corr} demonstrate a significant reduction of auto-correlation in the $L = 4\pi$ simulation compared to the $L = 2\pi$ simulation. The resulting Reynolds stress anisotropies are compared on the barycentric triangle in Fig. \ref{fig:auto-corr} and show that the two simulations yield nearly identical results. The effect of the remaining correlation is therefore taken to be small given these results.

Figure \ref{fig:refine} shows a convergence study demonstrating the effect of mesh refinement on Reynolds stress anisotropy for a forced simulation. This study is necessary since our LES methodology only uses the resolved portion of the Reynolds stress. Spatially and temporally averaged results are plotted on the barycentric triangle. Mesh refinement generally leads to an increase in isotropy of the Reynolds stress tensor, although variation in isotropy is limited for refinements beyond the $N=64^3$ mesh. We maintain this resolution for the remainder of this work as we expect diminishing returns by adding smaller scales to measure volume-averaged quantities. A procedure to account for the unresolved stresses and effectively increase the resolution of the simulations is detailed in Appendix \ref{app:SGS}, but this accounting for sub-grid scale (SGS) content was found to be unneeded at the chosen resolution.

\section{Stationary forced simulations}

\begin{figure}[b]
    \centering
    \includegraphics[width=\textwidth]{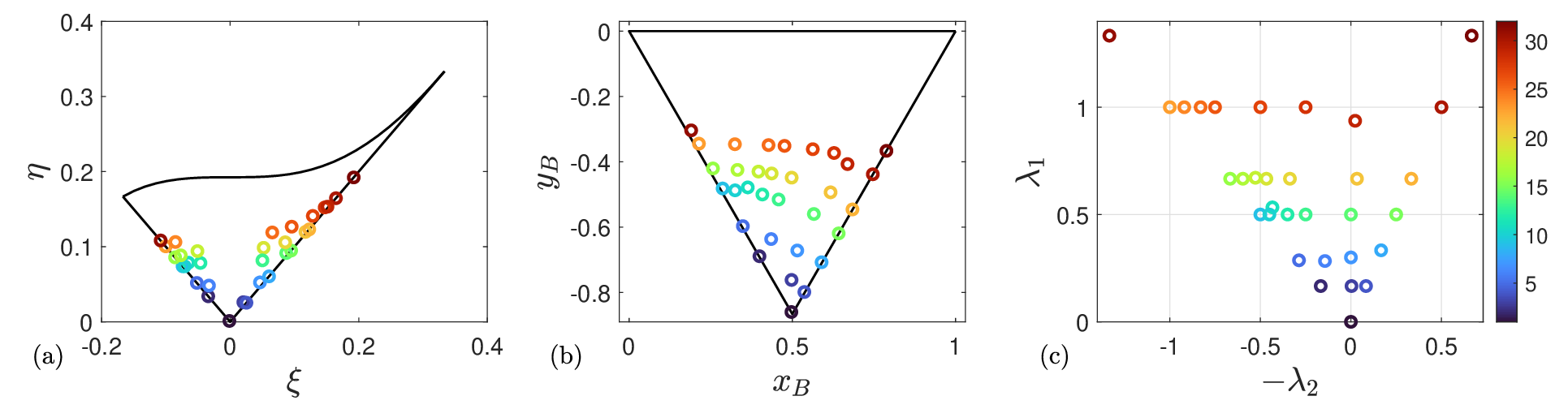}
    \caption{Stationary simulation results corresponding to the implemented diagonal forcing cases. Reynolds stress anisotropies are mapped on (a) the Lumley triangle and (b) the barycentric triangle. The first two eigenvalues of the trace-removed forcing matrix, $A_{\text{anis}} = A_{ij} - \frac{1}{3}A_{kk}\delta_{ij}$, are shown in (c), where the third eigenvalue is $-\lambda_1-\lambda_2$. Points are colored to allow reference between \textcolor{\commentcolor}{mappings. The color scale corresponds to the case numbers and forcing matrices of Table \ref{tab:allcases} in Appendix \ref{app:forcings}}.}
    \label{fig:forcing-cases}
\end{figure}

For the present anisotropically forced homogeneous turbulence, the RST equations become 

\begin{equation}
\frac{d{\overline{u_iu_j}}}{d{t}} = P_{ij} + D_{ij}
= 0
    \label{eqn:forced-RS}
\end{equation}

\noindent where $P_{ij} = \overline{u_i\Omega A_{jk}\Tilde{u_k}} + \overline{u_j\Omega A_{ik}\Tilde{u_k}}$ is the rate of turbulent energy production and $D_{ij} = \Pi_{ij} - \varepsilon_{ij}$ compactly represents the decay terms in Eqn. \ref{eqn:RS}. The previously-described forcing enables the system to be maintained at a stationary state such that the mean temporal term vanishes. A timeseries of the terms in the Reynolds stress budget for a forced simulation are given in Fig. \ref{fig:budget}. Plots shown are associated with the principal components of the Reynolds stress tensor. The pressure-strain term redistributes energy among the Reynolds stress components and therefore sums to zero across the principal components. The budget terms can be seen to closely balance each other, with a minor numerical residual ($O(1\%)$ of the TKE production term) primarily due to temporal fluctuations.

Specifications of the forcing matrix $A_{ij}$ were chosen to produce various Reynolds stress anisotropies. Forcing specifications were diagonal so that the results would be set in principal coordinates. Forced simulation results were post-processed to yield spatially and temporally averaged quantities. Independent samples were constructed by averaging over windows of $10 T$, following the guidance of \cite{shirian_2023}. Data corresponding to a transitional period of $10T$ were discarded from the beginning of each set of results to ensure decorrelation from the initial condition. A total of 49 samples were thus constructed for each simulation. The resulting Reynolds stresses were used to compute the normalized Reynolds stress anisotropy tensor.

Figure \ref{fig:forcing-cases} shows the resulting Reynolds stress anisotropies of 32 forced simulations on the anisotropy triangles. The specific values of these forcing matrices are listed with case numbers in Table \ref{tab:allcases} of Appendix \ref{app:forcings}. In Fig. \ref{fig:SSG-data}, we plot our simulations with other data that has been used to tune and assess RST decay models in \cite{chung_1995, sarkar_1990} and see that we consider a more diverse \textcolor{\commentcolor}{portfolio of} turbulent states than previous works. While the present methodology does not probe all regions of the triangle corresponding to extreme states of anisotropy, those regions are associated with near-wall flow, complex atmospheric conditions \cite{stiperski_2021}, or other states of turbulence that should not be expected to be present in wall-absent flows.

\begin{figure}[t]
    \centering
    \includegraphics[width=0.85\textwidth]{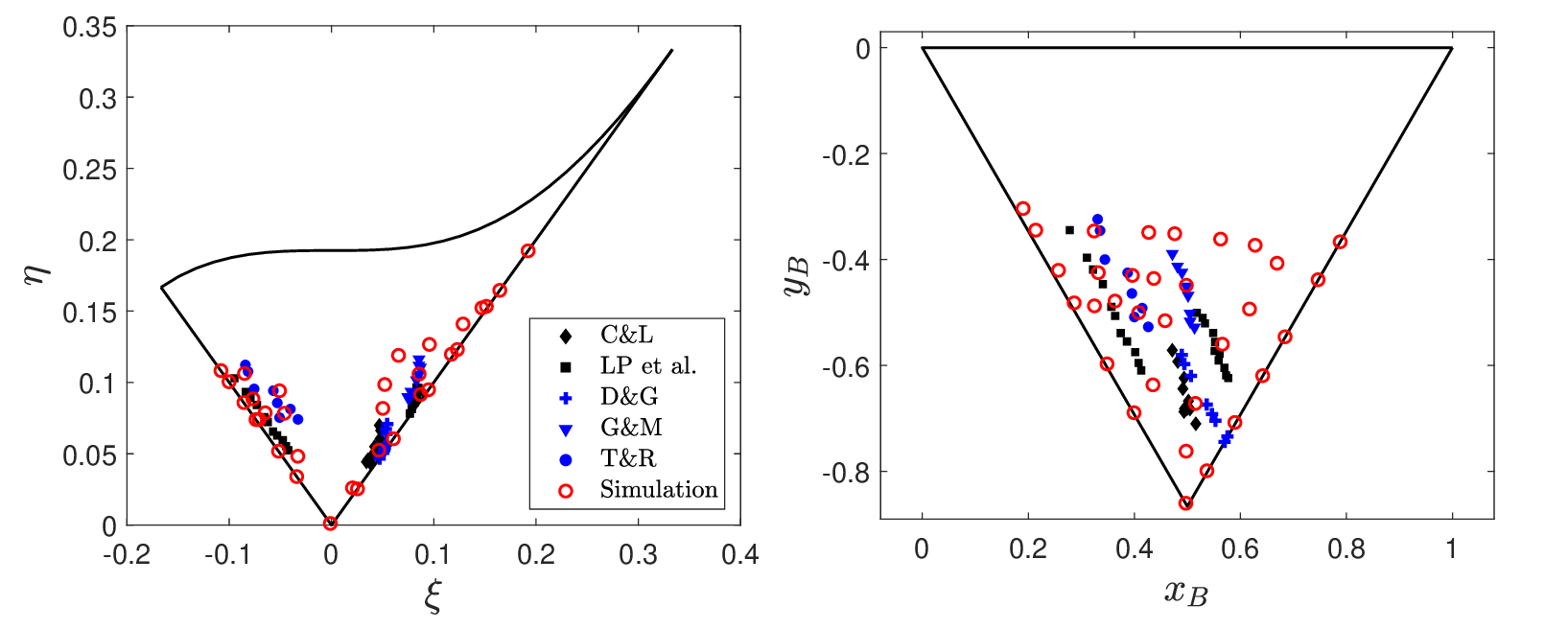}
    \caption{A comparison of the experimental data used to tune the models from \cite{sarkar_1990, chung_1995} and the HAT steady-state data used to tune the proposed cubic model on both invariant triangles. Experimental data comes from \cite{lepenven_1985, choi_1984, Gence_1980, reynolds_1968,dakao_1990}, with black symbols representing data used to tune the SSG decay term, blue the additional data shown in \cite{chung_1995}, and red denoting the original data of this work. Of note is that each red symbol represents a separate realization, whereas each set of black and blue symbols are measurements from single cases.}
    \label{fig:SSG-data}
\end{figure}

To characterize the range of turbulence states accessed using our forcing method, we can adopt a metric based on $F = 1 - 27\eta^2 + 54\xi^3$, which is the determinant of the normalized Reynolds stress tensor as listed in \cite{pope_2000}. As can be seen in Fig. \ref{fig:triangles}, $F = 0$ along the 1C--2C curve and monotonically increases to $F = 1$ at the $3\text{C}$ vertex of isotropic turbulence. As a result, the smallest value of $F$ that can be post-processed from our simulations allows us to quantify how close to filling in the entire space of Reynolds stress states we approach. Based on the cases in Appendix \ref{app:forcings}, the smallest value across each forcing matrix we report in this work is $F = 0.39$, achieved by a point on the 1C -- 3C leg.

\section{Model formulation}

In this section, we propose a modeling framework for capturing the decay term in HAT. Models like \cite{sarkar_1990, chung_1995} start by performing a Taylor-series expansion of $b_{ij}$ about the isotropic state ($b_{ij} = 0$) and then use the Cayley-Hamilton theorem to write $b^n_{ij}$ terms for $n > 2$ in terms of $b_{ij}$ and $b_{ik}b_{kj}$. This standard approach enforces certain physical constraints by construction, but does not preclude other forms. Instead of using such a formalism, we directly construct a general polynomial model form, which, through utilization of principal coordinates along with an appropriate selection of coefficients, satisfies rotational symmetry, zero-trace requirements, and realizability.

\subsection{Model form}

We consider a general decay model form that is a polynomial function of the normalized Reynolds stresses, $\tau_{ij} = \overline{u_iu_j}/2k$. This form has tunable constant coefficients and is structured to meet dimensional and orthogonality constraints. Note that $k$ is an explicitly known quantity in the RST framework. In principal coordinates, the model form for decay of the $\tau_{11}$ component is written as

\begin{equation}
\frac{d{\overline{u_1u_1}}}{d{t}} = D_{11} = -2\epsilon  f\left(\tau_{11},\tau_{22},\tau_{33}\right).
    \label{eqn:gen-model}
\end{equation}

While a timescale for the RST equations is formed by $\epsilon/k$, specifying a model for the dissipation, $\epsilon$, is beyond the scope of this work. We focus on the non-dimensional function $f$, which captures the anisotropy of the decay term. To this end, $f$ should be a symmetric function with respect to $\tau_{22}$ and $\tau_{33}$, given that $u_2$ and $u_3$ are orthogonal to $u_1$. Forms matching Eqn. \ref{eqn:gen-model} are also applied to the other two principal components. 

We find that defining $f\left(\tau_{11},\tau_{22},\tau_{33}\right)$ as a cubic polynomial of the Reynolds stresses significantly better predicts the forced simulation data than lower order polynomials. This aligns with the finding of \cite{Gence_1980, chung_1995} that Reynolds stress decay varies with the sign of the third invariant of $b_{ij}$, which is a cubic power of $b_{ij}$. Models involving higher-order selections of terms from the polynomial expansion for $f$ in Eqn. \ref{eqn:gen-model} were tuned and assessed, including up to the fourth powers of the Reynolds stresses and up to 12 terms. These did not show meaningful improvement over the cubic model, and some decreased performance for large models is likely attributable to a decrease in the effectiveness of the tuning procedure, as the number of free parameters increases relative to the number of data points available for least squares regression. A lower-polynomial model form including half powers of the Reynolds stresses was also considered as an alternative, but also did not convincingly offer improvements.

The proposed model cubic term is 

\begin{subequations}  \label{eqn:cubic-model}
\begin{align}
D_{11} = -2\epsilon \left(c_{1} \tau_{11} + c_{2}\left(\tau_{22} + \tau_{33}\right) + c_{3} \left(\tau_{11}\right)^2 + c_{4} \left(\left(\tau_{22}\right)^2 + \left(\tau_{33}\right)^2\right) + c_{5} \left(\tau_{11}\right)^3 + c_{6} \left(\left(\tau_{22}\right)^3 + \left(\tau_{33}\right)^3\right)\right),
    \label{eqn:cubic-model-x}
\\
D_{22} = -2\epsilon \left(c_{1} \tau_{22} + c_{2}\left(\tau_{11} + \tau_{33}\right) + c_{3} \left(\tau_{22}\right)^2 + c_{4} \left(\left(\tau_{11}\right)^2 + \left(\tau_{33}\right)^2\right) + c_{5} \left(\tau_{22}\right)^3 + c_{6} \left(\left(\tau_{11}\right)^3 + \left(\tau_{33}\right)^3\right)\right),
    \label{eqn:cubic-model-y}
\\
D_{33} = -2\epsilon \left(c_{1} \tau_{33} + c_{2}\left(\tau_{11} + \tau_{22}\right) + c_{3} \left(\tau_{33}\right)^2 + c_{4} \left(\left(\tau_{11}\right)^2 + \left(\tau_{22}\right)^2\right) + c_{5} \left(\tau_{33}\right)^3 + c_{6} \left(\left(\tau_{11}\right)^3 + \left(\tau_{22}\right)^3\right)\right),
    \label{eqn:cubic-model-z}
\end{align}
\end{subequations}

\noindent where $c_i$ are the model coefficients with final values given in Table \ref{tab:tuned-coef}.

Note that LRR decay term of Eqn. \ref{eqn:LR} can also be reformulated as a linear function of the principal components of the Reynolds stress tensor. For example, the equation governing $\overline{u_1u_1}$ is

\begin{equation}
\frac{d~\overline{u_1u_1}}{dt} = -2\epsilon \left( a_{1} \tau_{11} + a_{2} \left(\tau_{22} + \tau_{33}\right) \right),
\label{eqn:LR-reform}
\end{equation}

\noindent with coefficients $a_1 = 1.333$ and $a_2 = -0.167$ \textcolor{\commentcolor}{calculated from the original model parameter $C_1 = 1.5$ \cite{lrr_1975}}.  Similarly, the SSG model term of Eqn. \ref{eqn:SS} can be rewritten, with the evolution of $\overline{u_1u_1}$ dictated by

\begin{equation}
\frac{d~\overline{u_1u_1} }{dt} = -2\epsilon \left( b_{1} \tau_{11} + b_{2} \left(\tau_{22} + \tau_{33}\right) + b_{3} \left(\tau_{11}\right)^2 + b_{4} \left(\left(\tau_{22}\right)^2 + \left(\tau_{33}\right)^2\right) \right),
\label{eqn:SS-reform}
\end{equation}

\noindent with coefficients $b_1 = 2.4$, $b_2 = -0.7$, $b_3 = -1.4$, and $b_4 = 0.7$ \textcolor{\commentcolor}{calculated from the original model parameters $C_1 = 3.4$ and $C_2 = 4.2$ \cite{sarkar_1990}}.

\subsection{Constraints and realizability}

Before fitting to the data, we apply \textit{a priori} constraints to the six coefficients of the proposed model (Eqn. \ref{eqn:cubic-model}) to satisfy physical requirements. By definition $\frac{dk}{dt} = -\epsilon$, and so we require 

\begin{equation}
    D_{ii} = 2\frac{dk}{dt} =  -2\epsilon.
    \label{eqn:trace-decay}
\end{equation}

\noindent Given that \textcolor{\commentcolor}{the trace of $\tau_{ij}$ is $1$}, applying Eqn. \ref{eqn:trace-decay} to Eqns. \ref{eqn:cubic-model} leads to the following coefficient relations:

\begin{equation}
    c_1 + 2c_2  = 1, \quad \quad c_3 + 2c_4 = 0, \quad \quad c_5 + 2c_6 = 0.
\label{eqn:coef}
\end{equation}

Additionally, we want to ensure that the model yields realizable states of turbulence, as first proposed by \cite{schumann_1977}, but since clarified by many others, as reviewed in \cite{speziale_1994}. Following the arguments presented in 
\cite{sarkar_1990, speziale_1994}, we wish to ensure our model satisfies the \textit{weak} realizability condition, which essentially requires any trajectory given by the model to remain inside the bounded region of the invariant triangles, without allowing the turbulence to reach the one or two component limit states. Without loss of generality, this is equivalent to  

\begin{equation}
    \frac{d\overline{u_1u_1} }{dt} \geq 0 \quad \text{when} \quad \tau_{11} = 0.
\label{eqn:weak-real}
\end{equation}

Referring back to Eqn. \ref{eqn:cubic-model-x}, this neccesitates

\begin{equation*}
    c_{2} + c_{4} \left(\left(\tau_{22}\right)^2 + \left(1-\tau_{22}\right)^2\right) + c_{6} \left(\left(\tau_{22}\right)^3 + \left(1-\tau_{22}\right)^3\right) \leq 0,
\end{equation*}

\noindent or, 

\begin{equation} \label{eqn:realizability}
    \tau_{22}^2\left( 2c_4+3c_6 \right) + \tau_{22}\left( -2c_4 - 3c_6 \right) + \left( c_2 + c_4 + c_6 \right) \leq 0.
\end{equation}

This quadratic function on $\tau_{22} \in [0,1]$ has three potential extrema to check, which are the two boundaries and the global extremum at $\tau_{22} = 0.5$. If $2c_4+3c_6 \geq 0$, Eqn. \ref{eqn:realizability} has positive concavity on the interval and we require $ c_2 + c_4 + c_6 \leq 0$; otherwise, if $2c_4+3c_6 < 0$, Eqn. \ref{eqn:realizability} exhibits negative concavity and we require $ 4c_2+2c_4+c_6\leq 0$. 

\begin{figure}[t]
    \centering
    \includegraphics[width=0.85\textwidth]{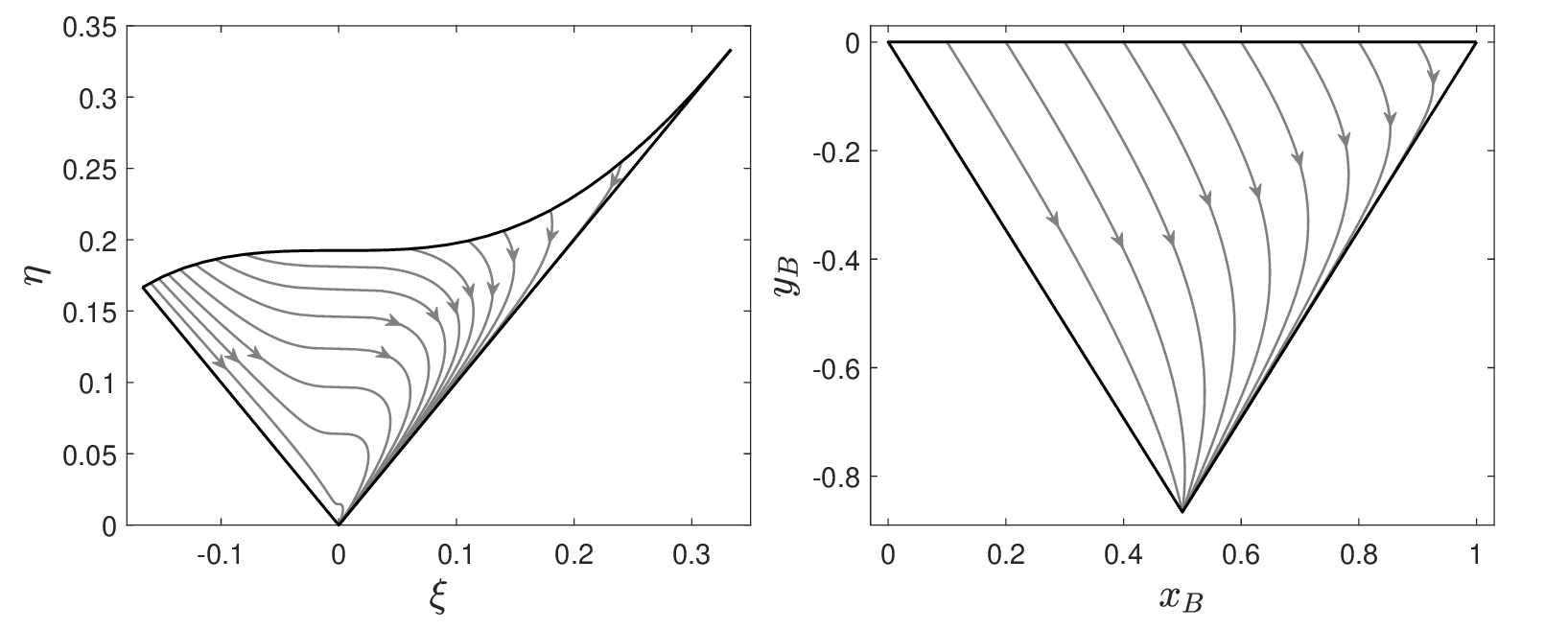}
    \caption{Cubic model decay trajectories calculated by solving Eqns. \ref{eqn:cubic-model} directly remain in realizable regions of the invariant triangles. Note the greater curvature of the streamlines in the Lumley triangle (L) compared to the barycentric map (R).}
    \label{fig:realizability}
\end{figure}

The strong version of the realizability condition would require $c_2 + c_4 + c_6 = 2c_4 + 3c_6 = 0$ and leave the proposed cubic decay term with only one model parameter. The strong realizability condition is essential only if the 1C and 2C states, corresponding to the upper boundary of the anisotropy triangles, are achieved. However, as we do not expect to access these extreme states of turbulence, it is overly restrictive as a constraint on model coefficients. It is sufficient, then, to satisfy the weak condition alone to generate a realizable decay model with three parameters that must satisfy certain constraints. An \textit{a postereriori} check of realizability is shown in Fig. \ref{fig:realizability}, where trajectories that start inside the anisotropy triangles remain inside the convex set of accessible states.

\subsection{Coefficient determination}

\begin{table}[b]
\large
\begin{tabular}{@{}llllll@{}}
$c_1$ &~~~~~~ $c_2$  &~~~~~~ $c_3$  &~~~~~~ $c_4$  &~~~~~~ $c_5$  &~~~~~~ $c_6$ \\ \midrule
 $8.2$ &~~~~~~ -$3.6$ &~~~~~~ -$14.9$ &~~~~~~ $7.5$ &~~~~~~ $11.3$ &~~~~~~ -$5.7$
\end{tabular}
\caption{The final cubic model coefficients, which satisfy realizability constraints and optimally fit the steady-state simulations in Fig. \ref{fig:steady}.}
\label{tab:tuned-coef}
\end{table}

The coefficient constraints of Eqns. \ref{eqn:coef} are applied to Eqns. \ref{eqn:cubic-model}, which are then substituted into Eqn. \ref{eqn:forced-RS} to yield a linear optimization problem with respect to the remaining free model coefficients. The quantities $\tau_{ij}$, $\epsilon$, and $P_{ij}$ are measured directly from the simulation data, allowing a set of three equations to be written for each forced simulation (\textit{i.e.}, one equation for each principal component of the Reynolds stress tensor) with all quantities known except for the coefficients. The simulation data therefore form a system of $3n$ equations, where $n$ is the total number of forced simulations and a weighted ordinary least squares was applied to this system to determine optimal coefficient values. Weights are determined by forming a Voronoi tessellation to find the proportion of the barycentric triangle represented by each flow realization, as seen in Fig. \ref{fig:voronoi}.

The resulting rounded coefficient values, which satisfy the constraints in the previous section, are provided in Table \ref{tab:tuned-coef}. Additionally, in Fig. \ref{fig:realizability} we see that the model produces trajectories that remain inside the invariant triangle, which supports the model's realizability. We further see that these trajectories are less curved in the barycentric triangle than in the Lumley triangle. The S-curve in \textcolor{\commentcolor}{the} Lumley triangle is a consequence of a highly nonlinear invariant mapping around the plane-strain $\xi=0$ line as opposed to a symptom of complex dynamical behaviour.

\section{Results and Discussion}

Using the presented framework for model form selection and coefficient determination, we first run simulations of steady, forced HAT to find optimal parameters. We then validate the model form against realizations of unforced decaying turbulence as they exhibit return-to-isotropy behavior, and finally offer some comments on the uniqueness of the $A_{ij}$ tensors probed.

\begin{figure}[b]
    \centering
    \includegraphics[width=0.85\textwidth]{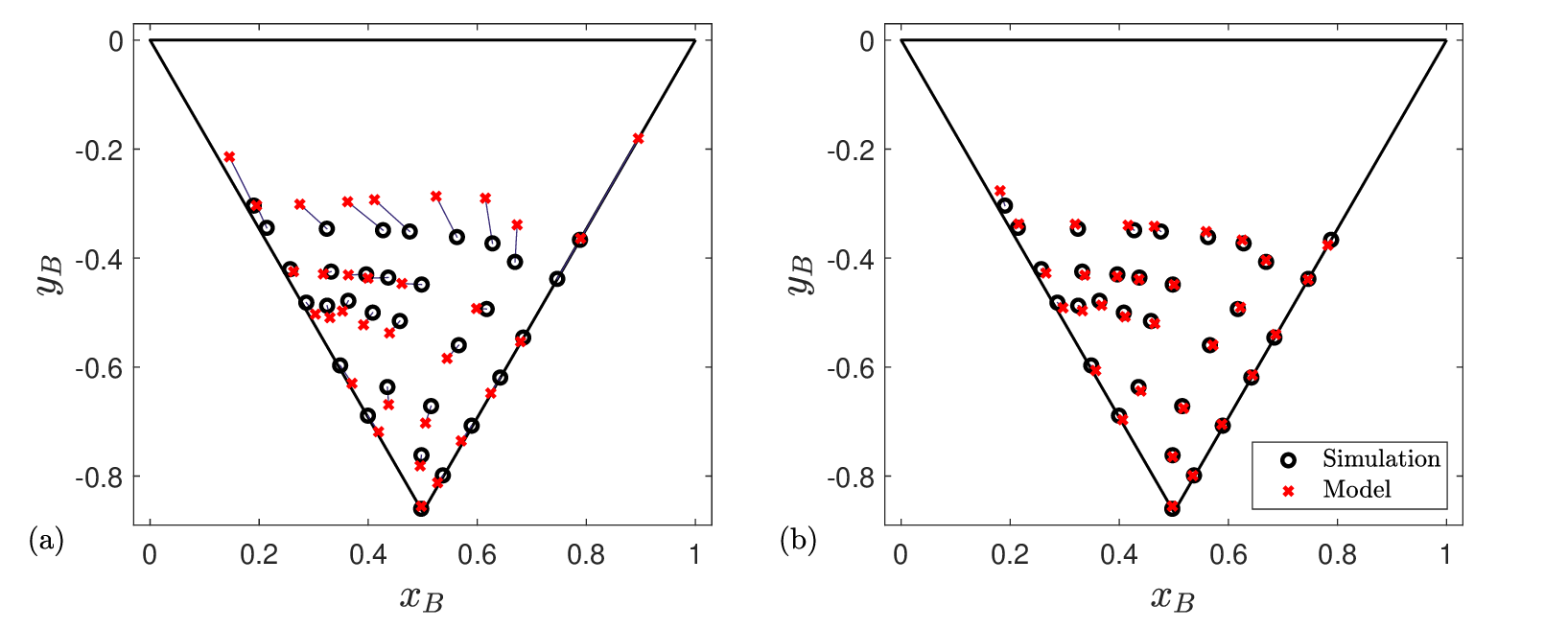}
    \caption{Stationary simulation results along with predictions from (a) the quadratic model form \textcolor{\commentcolor}{tuned using the present methodology} and (b) the cubic form tuned using the present methodology.}
    \label{fig:steady}
\end{figure}

\subsection{Stationary results and predictions} \label{sec:VA}

With the model coefficients determined using the described least squares procedure, we assess the accuracy of our fitting by numerically solving Eqn. \ref{eqn:RS}. We use the measured $P_{ij}$ from the simulation data and use our fitted models and the SSG model for the decay term $D_{ij}$. For each case, the resulting system of ordinary differential equations was solved to reach the steady state, and the resulting Reynolds stresses were compared against those obtained from the high-fidelity simulations. 

To properly assess the need for a cubic model form, we also fit model coefficients for Eqn. \ref{eqn:SS-reform} separately from those computed from the SSG model to show the best-possible quadratic model fit to the data. Akin to the SSG model, this quadratic model has only two free coefficients, constrained such that $b_1 + 2b_2  = 1, b_3 + 2b_4 = 0$ and with weak realizability requiring $b_4 \leq -2b_2$. This leads to $b_1 = 3.6, b_2 = -1.3, b_3 = -1.8, \text{ and } b_4=0.9$.

Fig. \ref{fig:steady} depicts the forced simulation results for the cubic model and the quadratic model, \textcolor{\commentcolor}{with coefficients $b$ and $c$ fitted using the data of this work and provided in this section, not the SSG or LRR methodologies}. While both models provide numerically stable solutions, the cubic model produces closer predictions for all of the data points. The quadratic model, on the other hand, is not able to as accurately predict anisotropies, especially far from the isotropic corner of the barycentric triangle.

Evaluating the performance of the two models in the stationary context, however, is complicated by the presence of the production term $P_{ij}$ in the stationary Reynolds stress equation, Eqn. \ref{eqn:forced-RS}. The production term, which is calculated from the steady simulation data, might introduce behavior associated with the forcing methodology that is not fully characteristic of naturally-occurring turbulence. For this reason, we next examine prediction of the developed model against unforced decaying turbulence data. The data of decaying simulations are not used in our model tuning process, therefore it allows not only assessment of the model, but also the methodology used for model fitting based on simulations of forced turbulence.  

\subsection{Decaying results and predictions}

\begin{figure}[t]
    \centering
    \includegraphics[width=0.75\textwidth]{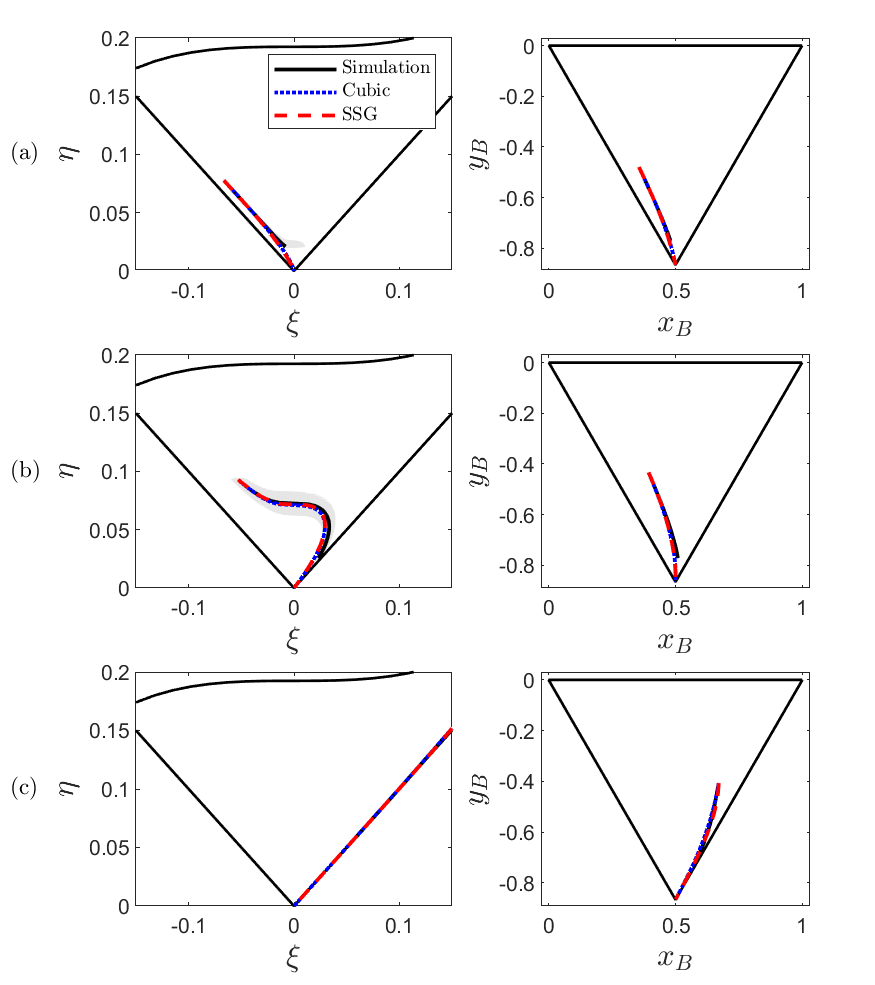}
    \caption{Turbulence decay trajectories and corresponding model predictions plotted on the Lumley triangle (left) and barycentric triangle (right) with initial conditions sampled from stationary simulations corresponding to forcing case 11 in row (a), 18 in row (b), and 29 in row (c). Shading indicates 95\% prediction intervals for the simulation data.}
    \label{fig:decay}
\end{figure}

\begin{figure}[ht]
    \centering
    \includegraphics[width=\textwidth]{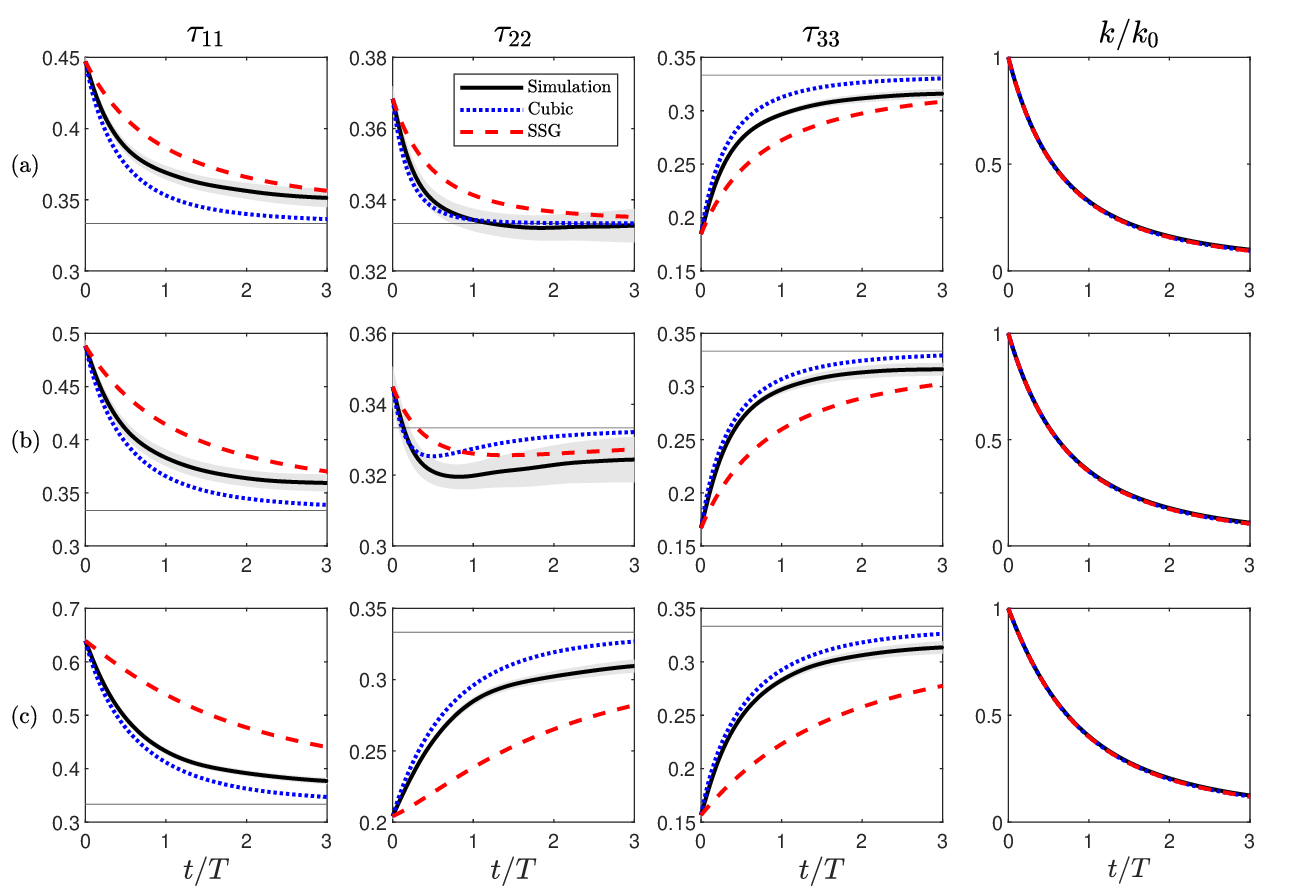}
    \caption{The evolution of simulation $\tau_{ij}$ values and corresponding model predictions are shown. The rows of the figure correspond to rows of Fig. \ref{fig:decay} with initial conditions sampled from stationary simulations corresponding to forcing case 11 in row (a); 18 in row (b); and 29 in row (c). The isotropy limit of $1/3$ is shown in a solid gray, with shading indicating 95\% confidence intervals for the simulation data.}
    \label{fig:timeplot}
\end{figure}

While stationary simulations are advantageous for model tuning as time-averaged data provides a convenient way of producing independent ensembles, unforced simulations of decaying turbulence assess model performance directly as the decay terms are the only active component of the RST equation

\begin{equation}
\frac{d{\overline{u_iu_j}}}{d{t}} = D_{ij}.
    \label{eqn:decaying-RS}
\end{equation}

Decay simulations are run by solving Eqn. \ref{eqn:momentum} without applied forcing. For each decay case, an ensemble of initial conditions are sampled from stationary simulation results corresponding to a given forcing case with sampling intervals of $5T$ to ensure independence of different ensembles. For each decay case, around 50 such simulations are performed using such spaced initial states, creating ensemble-averaged Reynolds stresses, which were then mapped to trajectories in the invariant triangles that can be statistically compared to model predictions.

Fig. \ref{fig:decay} shows several mean decay trajectories along with the corresponding model predictions. The uncertainty bounds in Fig. \ref{fig:decay} were generated using a statistical resampling procedure applied to the original decay ensembles of the Reynolds stress means at each time step. The computational bootstrap creates a sampling distribution for the Reynolds stresses from observed values \cite{james_2021}. From this distribution, samples were drawn and mapped to the anisotropy invariants $\xi$ and $\eta$. The 95\% prediction intervals were then determined for each invariant and used to construct ellipses around each point in the decay trajectory, which form the shaded regions in Fig. \ref{fig:decay} in aggregate.

The decay trajectories exhibit characteristics consistent with the findings of \cite{sarkar_1990, chung_1995, stiperski_2021} and others. Trajectories originating on or near the one- or two-component axisymmetric edges of the invariant triangle remain close to those respective lines as they approach the isotropic corner, while trajectories originating closer to the middle of the triangle tend to cross over toward the one-component axisymmetric side. Both SSG and the proposed cubic model demonstrate similar trajectories on both triangles and relatively close alignment with the simulation results.

Although such trajectories provide insight into the relative evolution of the Reynolds stresses, examining the complete dynamics of the system requires plotting that evolution as a function of time. This can make comparing models directly a complicated endeavor as models such as \cite{lrr_1975, ssg_1991, chung_1995} solve an auxiliary equation for $\epsilon$ to set a timescale independent of the data the rest of the model is fit to. While we do not propose an $\epsilon$ equation here, we have access to high-fidelity simulations and can instead measure $\epsilon(t)$ explicitly and directly provide it to the decay term models. In so doing, we guarantee that all models capture the TKE trajectory directly and any mispredictions are directly attributable to the model form. 

In Fig. \ref{fig:timeplot}, we show the evolution of the individual normalized Reynolds stress components for the high-fidelity simulations with predictions from the proposed cubic model and the standard SSG model. The proposed cubic model shows better agreement with the simulation data than the SSG model and better captures nonlinear behavior of the decay across the displayed cases. At large times, both SSG and the cubic model will asymptotically reach the isotropic condition of $\tau_{11}=\tau_{22}=\tau_{33} = \frac{1}{3}$, as physically expected. However, the timescale of this return-to-isotropy behavior is clearly better represented by the cubic model. We simulate to $t/T = 3$, which corresponds to the energy spectrum becoming strongly influenced by the domain size, as shown in Appendix \ref{app:filter-effects}. In Fig. \ref{fig:timeplot}, we see that the kinetic energy has also decayed significantly in this time interval. The simulation data in black is shown with 95\% confidence interval shading.

\subsection{The role of asymmetric forcing} \label{sec:asyforcing}

While we have thus far used diagonal -- and therefore symmetric -- forcing matrices to maintain principal coordinates in our stationary forced simulations, the effective forcing associated with mean velocity gradients in the production term of turbulent shear flows is typically asymmetric, as demonstrated in \cite{dhandapani_2019, rah_2018}. Our forcing methodology assumes that the decay of Reynolds stresses from a steady state condition is not substantially influenced by the precise nature of the production mechanism used to achieve that steady state condition. This implies that simulations with an asymmetrical $A_{ij}$ yield the same decay behavior as those with matched Reynolds stresses obtained from forcing with a diagonal $A_{ij}$ expressed in the principal coordinates of the Reynolds stress tensor. 

To test the validity of this assumption, we simulate an asymmetric forcing case with Reynolds stresses that match results from a diagonal $A_{ij}$ and examine the behavior of decaying turbulence initialized from those conditions. We choose an asymmetric forcing matrix corresponding to the one proposed by \cite{dhandapani_2019} as being analogous to nearly homogeneous shear turbulence, given as 

\begin{equation}
    \textcolor{\commentcolor}{\boldsymbol{A}_\text{Asymmetric}} = 
\begin{bmatrix*}[r]
0 & -1 & 0\\
0 & 0 & 0\\
0 & 0 & 0
\end{bmatrix*}.
    \label{eqn:A-asym}
\end{equation}

We then identify a diagonal \textcolor{\commentcolor}{symmetric} forcing matrix that matches the same principal component Reynolds stress as the asymmetrically forced simulation, given by 


\begin{figure}[b]
    \centering
    \includegraphics[width=\textwidth]{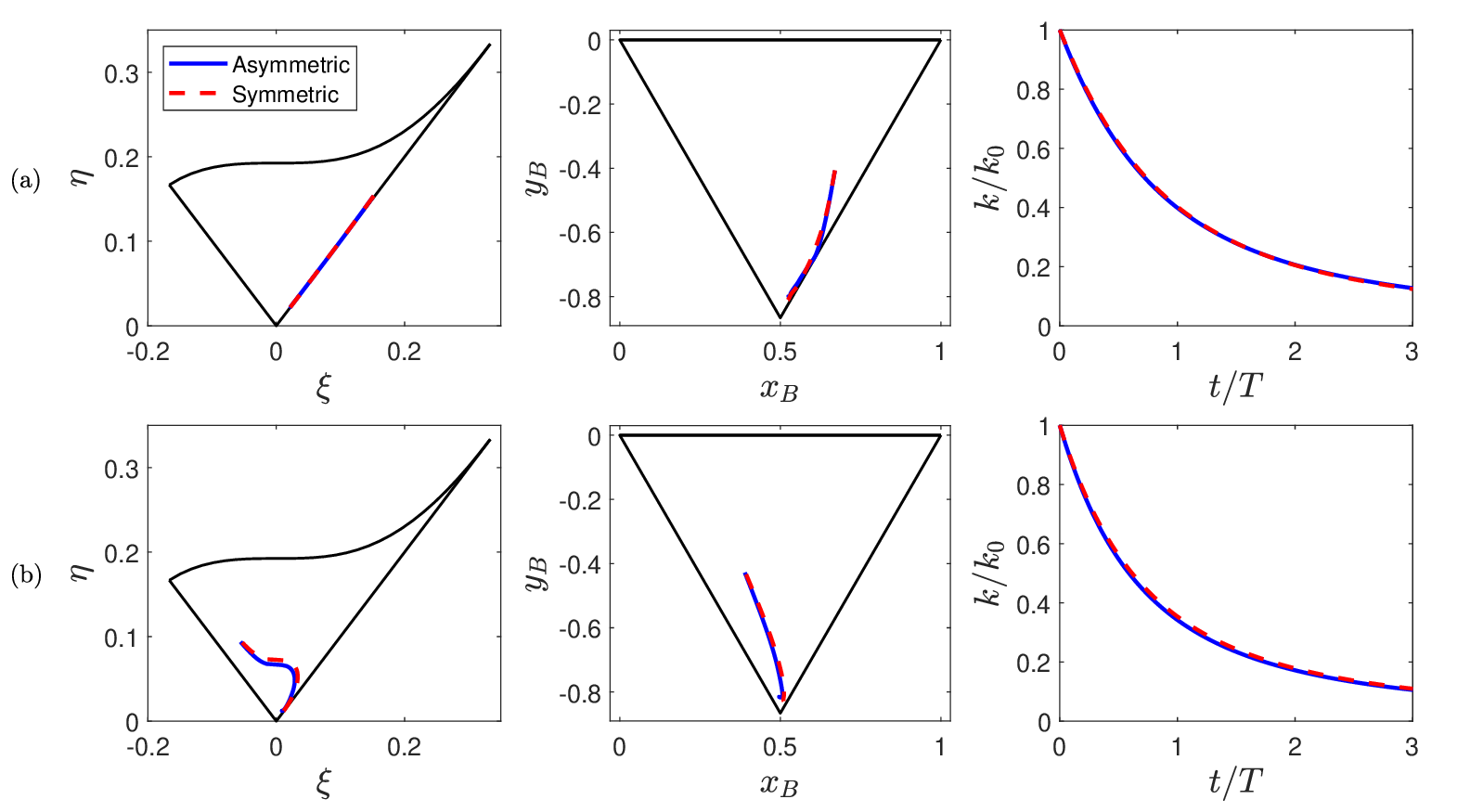}
    \caption{Decay from an initial condition generated by symmetric and asymmetric forcing with trajectories plotted on the Lumley triangle (left) and barycentric triangle (middle), as well as temporal decay (right). In row (a), $Ro=1$ for the asymmetric case (Eqn. \ref{eqn:A-asym}) and $Ro=\infty$ for the \textcolor{\commentcolor}{symmetric (diagonal)} case (Eqn. \ref{eqn:A-diag-c56}). In row (b), $Ro=1.9$ for the asymmetric case and $Ro=\infty$ for the \textcolor{\commentcolor}{symmetric (diagonal)} case.}
    \label{fig:asym}
\end{figure}

\begin{equation}
    \textcolor{\commentcolor}{\boldsymbol{A}_\text{Diagonal}} = 
\begin{bmatrix*}[r]
0.55 & 0 & 0\\
0 & -0.47 & 0\\
0 & 0 & 0.02
\end{bmatrix*}.
    \label{eqn:A-diag-c56}
\end{equation}

\begin{table}[t]
\begin{tabular}{llllllllll}
 &~~~~~~ $P_{11}$ &~~~~~~ $P_{22}$ &~~~~~~ $P_{33}$ \\ \hline
Asymmetric   &~~~~~~ 4.15 &~~~~~~ 0.00 &~~~~~~ -1.00
\\
\textcolor{\commentcolor}{Diagonal (symmetric)}    &~~~~~~ 3.66 &~~~~~~ 0.05 &~~~~~~ -0.74               
\end{tabular}
\caption{Reynolds stress budget term comparison in principal coordinates between the asymmetric forcing case (Eqn. \ref{eqn:A-asym}) and a corresponding \textcolor{\commentcolor}{diagonal (symmetric)} forcing case (Eqn. \ref{eqn:A-diag-c56}).}
\label{tab:budget-comp}
\end{table}

The resulting Reynolds stress budget terms from running a steady-state simulation with these two forcing matrices in Eqn. \ref{eqn:forced-RS} are given in Table \ref{tab:budget-comp}. The production terms, which Eqn. \ref{eqn:RS} shows are equivalent to the decay terms at steady state, have only slight differences between the two cases. Fig. \ref{fig:asym} shows two comparisons of decay from initial Reynolds stresses generated using symmetric and asymmetric forcings on the invariant triangles, along with decay of the TKE as a function of time. The decay trajectories appear nearly coincident, demonstrating that the assumptions on forcing hold at least as long as the anti-symmetric part of $\boldsymbol{A}$ is not larger than its symmetric part. We now examine a quantitative measure for this behavior.

We can return to the analogy relating the forcing matrix and mean velocity gradient tensor as covered in Sec. \ref{sec:gov-eqns}. Following this analogy, we can decompose the forcing matrix into symmetric and anti-symmetric parts as 

\begin{equation}
\boldsymbol{A} = \frac{1}{2}\left(\boldsymbol{A}+\boldsymbol{A}^T\right) + \frac{1}{2}\left(\boldsymbol{A}-\boldsymbol{A}^T\right) = \boldsymbol{A}_S + \boldsymbol{A}_{\Omega},
\end{equation}

\noindent where $\bullet^T$ denotes a transpose, \textcolor{\commentcolor}{while} $\boldsymbol{A}_S$ denotes the symmetric part of $\boldsymbol{A}$ and $\boldsymbol{A}_{\Omega}$ the anti-symmetric. To quantify the magnitude of the symmetric part relative to the anti-symmetric part, we can define a Rossby number $Ro \equiv \sqrt{\frac{A_{S,ij}A_{S,ij}}{A_{\Omega,ij}A_{\Omega,ij}}}$. In the limit of $Ro \rightarrow 0$, rotational effects become significant and can influence the evolution of the turbulence \cite{yeung_1998, baqui_2015}. Eqns. \ref{eqn:A-asym} and \ref{eqn:A-diag-c56} represent cases where $Ro = 1$ and $Ro = \infty$, respectively.

\begin{figure}
    \centering
    \includegraphics[width=0.5\textwidth]{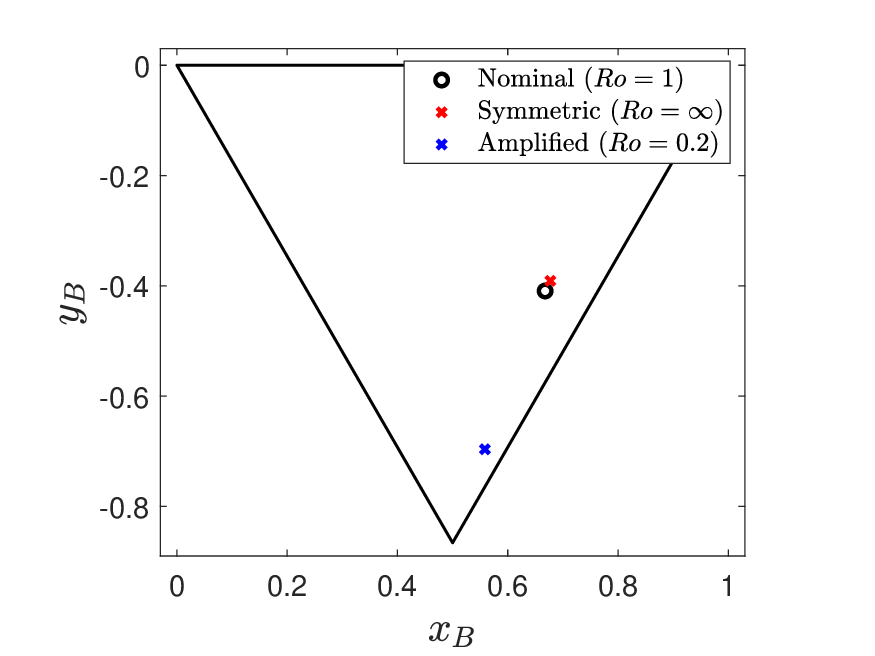}
    \caption{Reynolds stress anisotropies associated with the nominal asymmetric forcing matrix (Eqn. \ref{eqn:A-asym}), the symmetric part of the forcing matrix $\boldsymbol{A}_S$ (Eqn. \ref{eqn:A-OD_sym}), and the anti-symmetric part of the forcing matrix amplified (Eqn. \ref{eqn:A-c57}).}
    \label{fig:forcing-decomp}
\end{figure}

Simulations were performed corresponding to the asymmetric forcing case of Eqn. \ref{eqn:A-asym} with only the symmetric part of the forcing active so that

\begin{equation}
    \textcolor{\commentcolor}{\boldsymbol{A}_\text{S}}  = \frac{1}{2}
\begin{bmatrix*}[r]
0 & -1 & 0\\
-1 & 0 & 0\\
0 & 0 & 0
\end{bmatrix*}
    \label{eqn:A-OD_sym}
\end{equation}

\noindent are the forcing coefficients and $Ro = \infty$. \textcolor{\commentcolor}{Note that the anti-symmetric part of $\boldsymbol{A}_{Asymmetric}$ can be similarly defined.} Additional simulations were performed with the anti-symmetric part of Eqn. \ref{eqn:A-asym} amplified so that

\begin{equation}
    \textcolor{\commentcolor}{\boldsymbol{A}_\text{Amplified}} = \frac{1}{2}
\begin{bmatrix*}[r]
0 & -1 & 0\\
-1 & 0 & 0\\
0 & 0 & 0
\end{bmatrix*} + 
\frac{5}{2}
\begin{bmatrix*}[r]
0 & -1 & 0\\
1 & 0 & 0\\
0 & 0 & 0
\end{bmatrix*} = 
\begin{bmatrix*}[r]
0 & -3 & 0\\
2 & 0 & 0\\
0 & 0 & 0
\end{bmatrix*}, 
    \label{eqn:A-c57}
\end{equation}

\noindent and the results are shown in Fig. \ref{fig:forcing-decomp}, along with their $Ro$ values. Other tested cases showed similar results to those depicted.

The symmetric and asymmetric cases are closely located on the barycentric triangle, while the case with amplified anti-symmetry lies closer to the isotropic vertex. Thus, for a asymmetric forcing using our method where $Ro \gtrsim 1$, we expect that using purely the symmetric part of the forcing matrix yields similar results to use of the full asymmetric forcing. As many practical systems exist in this limit, this finding may be helpful in building confidence that our use of symmetric forcing in data generation has only minimally biased the resulting decay model and it can be used without modification in other scenarios.

\section{Conclusions} 

In this work, we sought to effectively inform modeling of the decay terms of the Reynolds stress transport equations. To combat the relative paucity of HAT data available, we leveraged the techniques of anisotropic forcing as potent tools to represent the essential anisotropy of turbulent flows in a computationally efficient manner. This work pragmatically built on linear forcing methods to generate realizations of anisotropic forcing that represent a portfolio of turbulent flows without involving boundary conditions. Use of a controller to prescribe the TKE value and wavenumber-dependent energy injection mitigated both statistical errors and finite domain size effects, while use of an explicit eddy viscosity model allowed Reynolds-number-independent tuning and assessment of an RST decay model.

A particular novelty of our modeling framework is the exclusive use of the principal components of the Reynolds stress tensor. The resulting model involves cubic terms of the Reynolds stress and is shown to provide improved performance to the SSG model in predicting the rate-of-change of the Reynolds stress components. Unlike previous models, we used a wide range of forced turbulence data to find model parameters, allowing us to use trajectories of decaying turbulence as a validation metric to assess model accuracy. 

As linear and quadratic models are commonly employed, the complexity of adding higher nonlinearities, like cubic terms, to the existing standard form may seem unnecessary. Indeed, we were able to fit model coefficients to only quadratic terms in Section \ref{sec:VA}, and the proposed cubic model trajectories in the invariant triangle do not appear to visually differ in a significant way from previous models. By disambiguating the effects of $\epsilon$ from the effects of $D_{ij}$ and examining the Reynolds stresses as a function of time, however, we demonstrated that the cubic model form captures the relevant physics to higher fidelity. This builds confidence that our improved results are not due to a fortuitous cancellation of errors, but are consistent with the recommendation of \cite{sarkar_1990} that additional nonlinearities should be explored as an avenue of improvement on their quadratic model and the work of \cite{chung_1995} in identifying the need for a cubic model form.

Future work will involve further investigation of model forms, higher-order nonlinearities, and wall effects that traditional redistribution models can often mispredict \cite{durbin_2018}. In addition, a key step towards a full RST model is an explicit representation of the function governing the dissipation rate, $\epsilon$, that remains unspecified in the present work. More future work could include the extension to lower Reynolds number conditions using corrections as in \cite{chung_1995}.

\section*{Acknowledgments} 
\noindent This work was supported by Office of Naval Research grant N000142212323 and the National Aeronautics and Space Administration under grant 80NSSC23M0225. T.H. and O.B.S. were supported in part by National Science Foundation Graduate Research Fellowship Program grants DGE-1656518 and DGE-2146755 and the Stanford Graduate Fellowships in Science and Engineering. The authors gratefully acknowledge contributions by Y.R. Yi, D.L.O.-L. Lavacot, and H. Le. 

\appendix 

\section{Effects of energy injection} \label{app:filter-effects}

\begin{figure}[t]
    \centering
    \includegraphics[width=0.5\textwidth]{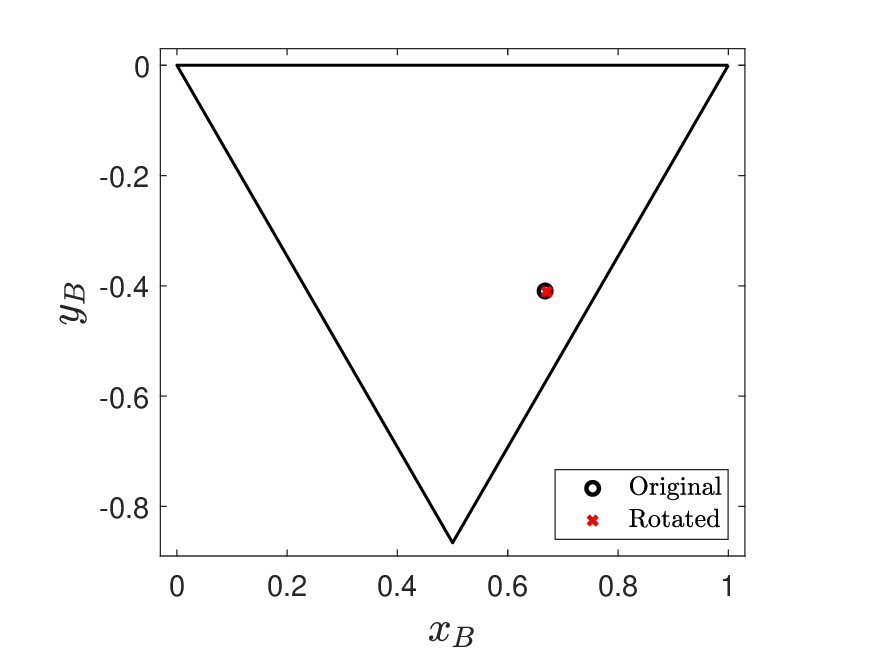}
    \caption{Comparison of Reynolds stress anisotropies corresponding to the asymmetric forcing matrix of Eqn. \ref{eqn:A-asym} with and without an applied $\theta = \pi/4~\vec{e}_3$ rotation.}
    \label{fig:rotation}
\end{figure}

The goal of the filtered energy injection method is to enforce a length-scale on the flow independent of the domain size itself. This enforcement will have implications for both the steady and decaying simulations.

\begin{figure}[b]
    \centering
    \includegraphics[width=0.5\textwidth]{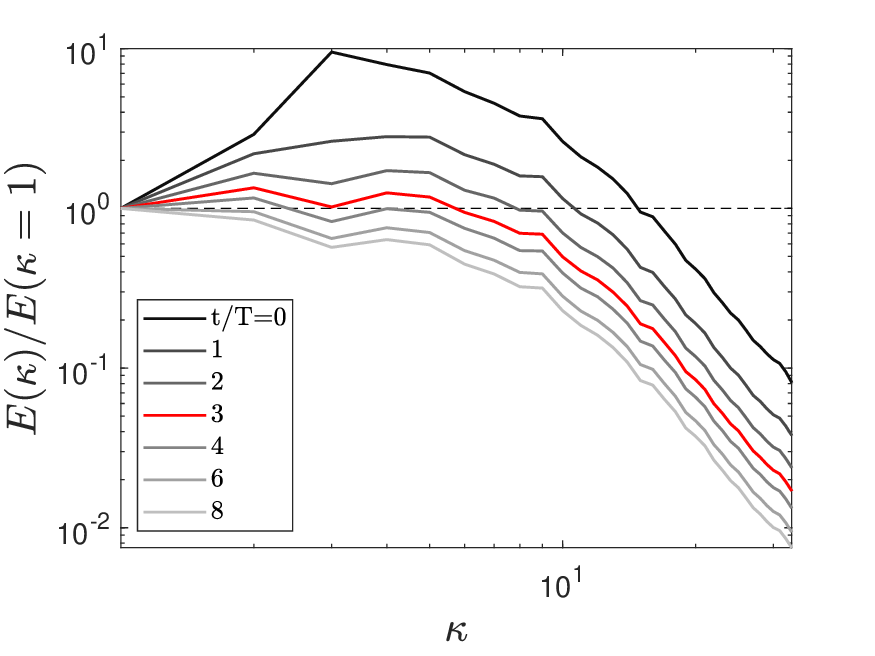}
    \caption{Decaying energy spectrum starting from the forced condition associated with case $18$. We select $t/T=3$ as the threshold past which domain size begins to have a prominent effect.}
    \label{fig:spectrum}
\end{figure}

To verify the efficacy of the filtering strategy, we can examine the effects of simulation box orientation on Reynolds stress results by using a forcing matrix for a stationary simulation that is rotated relative to the diagonal formulation such that the principal axes of the Reynolds stresses are not aligned with the box faces. We apply a $\theta = \pi/4~\vec{e}_3$ rotation to the asymmetric forcing matrix (Eqn. \ref{eqn:A-asym}) in Sec. \ref{sec:asyforcing} to get a newly-rotated forcing matrix of 

\begin{equation}
A_{ij} = 
\begin{bmatrix*}[r]
-0.4366 & -0.2563 & 0\\
0.7437 & 0.4366 & 0\\
0 & 0 & 0
\end{bmatrix*}.
\label{eqn:A-rotated}
\end{equation}

We run a turbulence simulation to a statistically stationary state with this forcing matrix and in Fig. \ref{fig:rotation}, we compare the Reynolds stress anisotropies from the original and rotated simulations. We see that they match well, demonstrating that box orientation effects have been minimized.

In addition to the forced simulations, we examine how the domain size influences decaying simulations once energy injection ceases. For these cases, we can examine the three-dimensional energy spectrum and judge when the domain size effects become relevant. In a $L = 2\pi$ box, this corresponds to the energy spectrum peak shifting to the smallest wavenumber of $k = 1$. In Fig. \ref{fig:spectrum}, we plot the three-dimensional energy spectrum in a box corresponding to case $18$ in Table \ref{tab:allcases} that starts at the forced condition at $t/T = 0$. We can see that this shift occurs around $t/T = 3$, which marks the end of our observation of the initial decay.

\section{\textcolor{\commentcolor}{Resolution effects and} sub-grid scale content} \label{app:SGS}

\begin{table}
\color{\commentcolor}
\begin{tblr}{cells ={c},row{even}={Light},vline{2}={-}{},}
N &~~~ $P_{11}$ &~~~ $P_{22}$ &~~~ $P_{33}$ &~~~ $2\epsilon$ &~~~ $\overline{u_1u_1}$ &~~~ $\overline{u_2u_2}$ &~~~ $\overline{u_3u_3}$ \\ \hline
$32^3$   &~~~ 3.61 (0.81) &~~~ 0.82 (0.19) &~~~ 0 &~~~ 4.46 &~~~ 1.65 (0.55) &~~~ 0.75 (0.25) &~~~ 0.58 (0.19) \\
$64^3$   &~~~ 3.49 (0.79) &~~~ 0.92 (0.21) &~~~ 0 &~~~ 4.43 &~~~ 1.54 (0.52) &~~~ 0.81 (0.27) &~~~ 0.64 (0.21) \\
$128^3$    &~~~ 3.41 (0.79) &~~~ 0.90 (0.21) &~~~ 0 &~~~ 4.32 &~~~ 1.53 (0.51) &~~~ 0.80 (0.27) &~~~ 0.66 (0.22) \\
$256^3$    &~~~ 3.22 (0.79) &~~~ 0.85 (0.21) &~~~ 0 &~~~ 4.07 &~~~ 1.53 (0.51) &~~~ 0.80 (0.27) &~~~ 0.67 (0.22)
\end{tblr}
\caption{Effects of grid refinement on Reynolds stresses and production total mean estimates along with dissipation values for case 14. Trace-normalized values are in parentheses and show monotone convergence by the $N=64^3$ case for both production and Reynolds stresses.}
\label{tab:production_convergence}
\end{table}

While we have shown explicit convergence of our method with grid resolution, we describe a method here to assess the effects of unresolved SGS turbulence in the model tuning and evaluation framework. The present large-eddy simulations probe the limit state of infinite Reynolds number turbulence by incorporating the diffusive effects of unresolved scales via an eddy viscosity in Eqn. \ref{eqn:momentum}. However, the Reynolds stresses used in the framework presented in Sec. IV are computed using only resolved fluctuations. Additionally, the production term $P_{ij}$ employed in the model tuning procedure should account for the forcing of unresolved spectral content as the grid imposes a filter scale that is smaller than that imposed by the high-pass filtered velocity field from the forcing term of Eqn. \ref{eqn:momentum}.

\begin{figure}[t]
    \centering
    \includegraphics[width=0.5\textwidth]{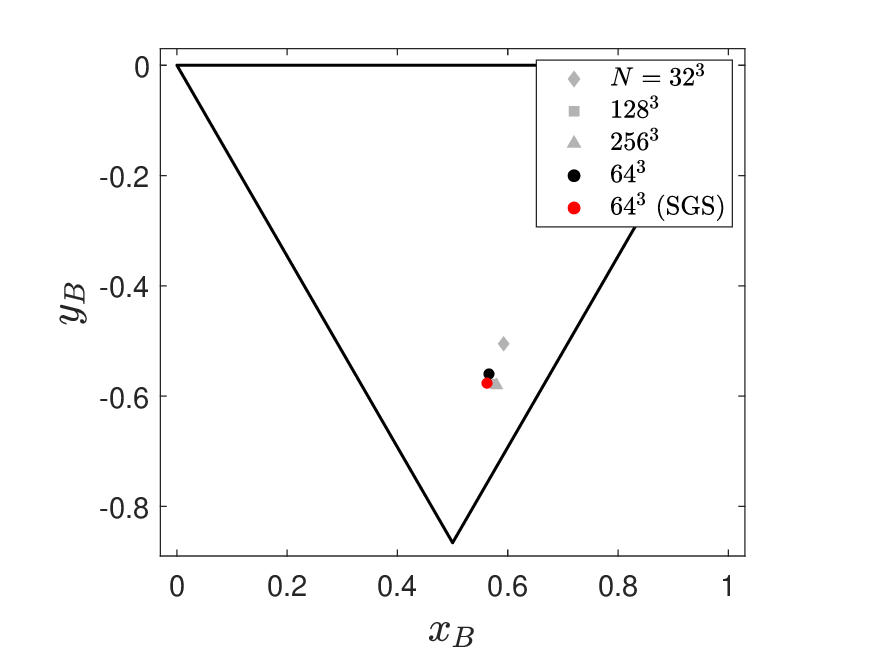}
    \caption{The convergence of Reynolds stress anisotropies with grid resolution on the barycentric triangle for case 14, including SGS stress estimation for the $N=64^3$ case.}
    \label{fig:sgs-est}
\end{figure}

\textcolor{\commentcolor}{To quantitatively assess resolution effects, the measured Reynolds stresses and turbulent production values are reported in Table \ref{tab:production_convergence} for case 14, for which anisotropy is significant. For $N \geq 64^3$, the Reynolds stresses are relatively invariant to the resolution and Figs. \ref{fig:refine} and \ref{fig:sgs-est} show that their anisotropies do not change significantly either. While the raw production values show greater sensitivity to the resolution, these values normalized by the trace of the production tensor, reported parenthetically, appear to reach grid convergence for $N \geq 64^3$.} 

\textcolor{\commentcolor}{Additionally, in equilibrium turbulence, the trace of the measured production tensor should balance the measured kinetic energy dissipation, and this is seen in Table \ref{tab:production_convergence}. As the current framework does not model the dissipation term and instead uses the measured value directly (\emph{e.g.} Eqns. \ref{eqn:forced-RS} and \ref{eqn:gen-model}), the model fitting is sensitive only to the relative anisotropy of the production and Reynolds stress components. Therefore, as the normalized production and Reynolds stress components are converged by $N=64^3$, this builds confidence that the resolution choices in this work are justified and SGS augmentation is unneeded. However, for completeness, we now outline a systematic procedure for how SGS effects could be accounted for.}

We can use an \emph{a posteriori} estimation procedure to compute the unresolved Reynolds stresses, denoted as $\overline{u_iu_j}^{\text{SGS}}$. While the Smagorinsky-Lilly model used to capture sub-grid effects in our simulations computes the deviatoric part of the sub-grid scale stress tensor as

\begin{equation}     \label{eqn:dev-SGS}
        \overline{u_iu_j}^{\text{SGS}} - \frac{1}{3}\overline{u_ku_k}^{\text{SGS}}\delta_{ij} = -2\left(C_s\Delta\right)^2|\overline{S}|S_{ij},
\end{equation}

\noindent we can compute the trace $\overline{u_ku_k}^{\text{SGS}}$ following \cite{yoshizawa_1985} as

\begin{equation}     
\label{eqn:tr-SGS}
    \overline{u_ku_k}^{\text{SGS}} = 2\left(C_I\Delta\right)^2|\overline{S}|\textcolor{\commentcolor}{^2},
\end{equation}

\noindent where $C_I = \textcolor{\commentcolor}{0.164}$ was tuned to match the highest resolution simulated. The resulting SGS stress tensor is then added to the resolved Reynolds stress tensor and shown alongside simulation results associated with various grid resolutions in the anisotropy plot in Fig. \ref{fig:sgs-est} and with numerical values in \textcolor{\commentcolor}{Table \ref{tab:SGS_correction}}. We can see that the SGS-added point lies close to the \textcolor{\commentcolor}{existing} simulation results and that the Reynolds stresses are relatively invariant to the resolution for $N \geq 64^3$.  \textcolor{\commentcolors}{As an aside, for $N = 64^3$, we can now calculate $M = \frac{k_\text{SGS}}{k_\text{Resolved} + k_\text{SGS}} = 1 - \frac{k_\text{Resolved}}{k_\text{Resolved}+k_\text{SGS}} = 0.05$ as a metric of LES resolution, inspired by Eqn. 1 in \textcite{pope_2004}. As $M = 0$ corresponds to DNS and $M = 1$ to RANS, it is reasonable to conclude our LES is sufficiently resolved.}

\begin{table}
\color{\commentcolor}
\begin{tblr}{cells ={c},row{even}={Light},vline{2}={-}{},}
N &~~~ $P_{11}$ &~~~ $P_{22}$ &~~~ $P_{33}$ &~~~ $\overline{u_1u_1}$ &~~~ $\overline{u_2u_2}$ &~~~ $\overline{u_3u_3}$ \\ \hline
$64^3$   &~~~ 3.49 (0.79) &~~~ 0.92 (0.21) &~~~ 0 &~~~ 1.54 (0.52) &~~~ 0.81 (0.27) &~~~ 0.64 (0.21) \\
\emph{$64^3$ + SGS (raw)}   &~~~ \emph{3.61 (0.79)} &~~~ \emph{0.98 (0.21)} &~~~ \emph{0} &~~~ \emph{1.59 (0.50)} &~~~ \emph{0.86 (0.27)} &~~~ \emph{0.70 (0.22)} \\
$64^3$ + SGS (norm.)  &~~~ 3.44 (0.79) &~~~ 0.94 (0.21) &~~~ 0 &~~~ 1.51 (0.50) &~~~ 0.82 (0.27) &~~~ 0.67 (0.22) \\
$128^3$    &~~~ 3.41 (0.79) &~~~ 0.90 (0.21) &~~~ 0 &~~~ 1.53 (0.51) &~~~ 0.80 (0.27) &~~~ 0.66 (0.22) \\
\end{tblr}
\caption{Effects of grid refinement on \textcolor{\commentcolors}{resolved} Reynolds stresses and production total mean estimates with dissipation values for case 14 and trace-normalized values are in parentheses. The raw SGS estimation for $N=64^3$ adds to both the production and Reynolds stresses. The normalized (norm.) row divides the production and Reynolds stresses such that $\overline{u_iu_i} = 3.0$.}
\label{tab:SGS_correction}
\end{table}

While not accounted for in our model tuning in this paper, the production tensor can also be consistently modified to reflect forcing of the unresolved scales. The unresolved content contribution to turbulent production is \textcolor{\commentcolor}{modeled as}

\begin{equation}     
\label{eqn:P-SGS}
        P_{ij}^{\text{SGS}} = \overline{\Omega A_{ik}\overline{u_ku_j}^{\text{SGS}}},
\end{equation}

\noindent which can be added to the resolved production values to change model tuning. \textcolor{\commentcolor}{Table \ref{tab:SGS_correction} shows the effects of the SGS estimation procedure and in raw terms, the sum of the resolved and SGS TKE is larger than the controller-set value. As Eqn. \ref{eqn:P-SGS} shows that $P_{ii}$ is proportional to the SGS TKE value, the simplest adjustment is to divide both the production and Reynolds stresses by the value needed to match the TKE setpoint to get the \emph{normalized} table row. We see this SGS estimation ansatz successfully adjusts $N = 64^3$ production data \textcolor{\commentcolors}{towards} the $N = 128^3$ data.}

\textcolor{\commentcolor}{As a final note, we address the non-convergence of production values in Table \ref{tab:production_convergence}. We assess that the monotonically decreasing production values do not arise from statistical uncertainty as our measured $95\%$ confidence interval for the production is at most $\sim \pm 2\%$ of the mean estimate for the most uncertain $N=256^3$ case. In contrast, \cite{shirian_2022} reported a 95\% confidence interval for mean dissipation over $O(500)$ eddy turnover times as $\sim \pm 6\%$ the mean estimate for DNS of stationary, linearly-forced homogeneous isotropic turbulence. Our use of controller-based forcing, therefore, measures turbulence statistics very reliably over shorter time horizons. We can therefore hypothesize that the decreasing production is an inherent consequence of the controller maintaining constant resolved TKE in lieu of a constant energy injection rate. The former forces the Reynolds stresses to quickly asymptote, while the latter would more directly impact the turbulent energy production term.}

However, our assessment of the sensitivity of the Reynolds stresses to variations of production in Table \ref{tab:production_convergence} suggest that the error in capturing production for $N=64^3$ simulation is of the order of the error committed by the fitting procedure, as seen in Fig. \ref{fig:steady}. We therefore do not expect resolution effects for the production values to affect our conclusions. 


\section{List of forcing matrices used} \label{app:forcings}

\begin{figure}[t]
    \centering
\includegraphics[width=0.5\textwidth]{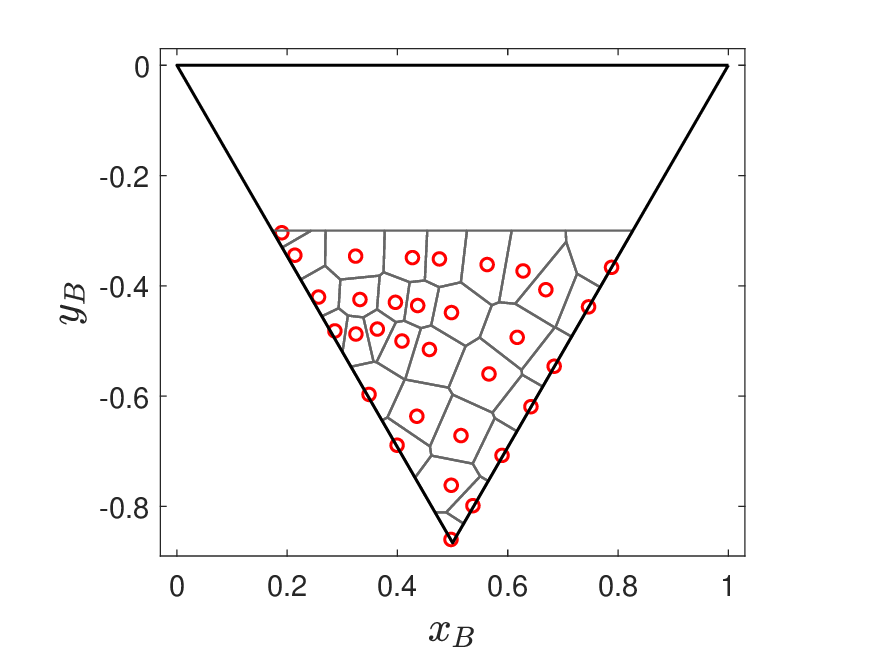}
    \caption{Stationary simulation results shown with the Voronoi tessellation used to assign weights in the model tuning procedure. The upper boundary corresponds to $y_B = -0.3$.}
    \label{fig:voronoi}
\end{figure}

Here, we list the diagonal, symmetric forcing matrices corresponding to the anisotropic cases used in this work: the asymmetric cases used are detailed in the relevant section. In Fig. \ref{fig:voronoi}, we can see the Voronoi tessellation on the barycentric map used to weight points during the linear fitting procedure. Placing an upper boundary for the Voronoi tessellation at $y_B = -0.3$ avoided over-weighting the highest row of data points on the barycentric triangle. However, model tuning results were found to be insensitive to the exact choice of this upper boundary.

To interpret all the case data, consider the forcing matrix written as

\begin{equation} 
A_{ij} = 
\begin{bmatrix}
A_{11} & 0 & 0\\
0 & A_{22} & 0\\
0 & 0 & A_{33}
\end{bmatrix},
\end{equation}

\noindent and see the data presented in Table \ref{tab:allcases}. The cases shown in Fig. \ref{fig:decay} \textcolor{\commentcolor}{and in Fig. \ref{fig:timeplot}} correspond to numbers 11, 18, and 29. \textcolor{\commentcolor}{Cases are colored in Fig. \ref{fig:forcing-cases} by case numbers, showing that cases are roughly ordered by increasing anisotropy of the Reynolds stresses.}

\begin{table} 
\begin{tblr}{cells ={c},row{even}={Light},column{3,9}={Dark},column{5,6}={White},row{1}={White},vline{2,8} = {-}{},}
Case & $A_{11}$ & $A_{22}$ & $A_{33}$ & \quad \quad & \quad \quad & Case & $A_{11}$ & $A_{22}$ & $A_{33}$ \\
1    & 1        & 1        & 1         &&& 17   & 1        & 0.93     & -0.93     \\
2    & 1        & 1        & 0.5       &&& 18   & 1        & 0.856    & -0.87     \\
3    & 1        & 0.83     & 0.67      &&& 19   & 1        & 0.8      & -0.8      \\
4    & 1        & 0.75     & 0.75      &&& 20   & 1        & 0.67     & -0.67     \\
5    & 1        & 1        & 0.14      &&& 21   & 1        & 0.3      & -0.3      \\
6    & 1        & 0.86     & 0.29      &&& 22   & 1         & 0        & 0        \\
7    & 1        & 0.7      & 0.4       &&& 23   & 1        & 1        & -2        \\
8    & 1        & 0.5      & 0.5       &&& 24   & 1        & 0.92     & -1.92     \\
9    & 1        & 1        & -0.5      &&& 25   & 1        & 0.83     & -1.83     \\
10   & 1        & 0.95     & -0.45     &&& 26   & 1        & 0.75     & -1.75     \\
11   & 1        & 0.9      & -0.5      &&& 27   & 1        & 0.5      & -1.5      \\
12   & 1        & 0.85     & -0.35     &&& 28   & 1        & 0.25     & -1.25     \\
13   & 1        & 0.75     & -0.25     &&& 29   & 1        & 0.04     & -0.85     \\
14   & 1        & 0.5       & 0        &&& 30   & 1        & -0.5     & -0.5      \\
15   & 1        & 0.25     & 0.25      &&& 31   & 1        & 1        & -3        \\
16   & 1        & 1        & -1        &&& 32   & 1        & -1       & -1         
\end{tblr}
\caption{Forcing matrix values for all diagonal cases used for fitting the cubic model of Eqns. \ref{eqn:cubic-model}. \label{tab:allcases}}
\end{table}

\bibliography{references}

\end{document}